\documentclass{mn2e}

\usepackage{graphicx}
\usepackage{amsmath}
\usepackage{url}

\title[Particles accelerated in shells of Novae]
{For how long are particles accelerated in shells of recurrent Novae ?}
\author[W. Bednarek]
{
W. Bednarek 
\\ 
University of \L \'od\'z, Department of Astrophysics, Faculty of Physics and Applied Informatics,
ul. Pomorska 149/153, 90-236 \L \'od\'z, Poland,
\\
wlodzimierz.bednarek@uni.lodz.pl
\\}
\begin{document}

\date{Accepted . Received ; in original form }

\pagerange{\pageref{firstpage}--\pageref{lastpage}} \pubyear{2015}

\maketitle

\label{firstpage}
\begin{abstract}
Galactic Novae is at present well established class of $\gamma$-ray sources. We wonder for how long the mechanism of 
acceleration of electrons operates in shells of Novae. In order to put constraints on the time scale of the
electron acceleration, we consider a specific model for the injection and propagation of electrons within
the shell of the recurrent Nova RS Ophiuchi. We calculate the equilibrium spectra of electrons within the Nova shell
and the $\gamma$-ray fluxes produced by these electrons in the comptonization of the soft radiation from the Red Giant within the Nova binary system and also radiation from the Nova photosphere. We investigate two component, time 
dependent model in which a spherically ejected Nova shell propagates freely in the polar region of the Nova binary 
system. But, the shell is significantly decelerated in the dense equatorial region of the binary system. We discuss 
the conditions for which electrons can produce $\gamma$-rays which might be detectable by the present and/or future 
$\gamma$-ray observatories. It is concluded that freely expending shells of Novae in the optimal case (strongly magnetised shell and efficiency of acceleration of electrons of the order 
of 10$\%$) can produce TeV $\gamma$-rays within the sensitivity of the Cherenkov Telescpe Array only within 1-2 years after explosion.  On the other hand, decelerated shells of Novae have a chance to be detected 
during the whole recurrence period of RS Ophiuchi, i.e. $\sim$15 years.
\end{abstract}
\begin{keywords} 
stars: novae --- acceleration of particles --- radiation mechanisms: non-thermal --- gamma-rays: stars 
--- cosmic rays
\end{keywords}

\section{Introduction}

Nova is a thermonuclear explosion in a layer of matter on the surface of a White Dwarf (WD). 
The matter appears as a result of the accretion process from a companion star in the WD binary system 
(Gallagher \& Starrfield~1978, Bode \& Evans~2008). 
Depending on the type of the companion star, a Red Giant (RG) or a Main Sequence star, the Nova is called 
a Symbiotic or a Classical Nova. Explosions occur on different time 
scales in different binary systems, depending on the mass of the WD and the accretion rate. 
Novae, observed many times in a human life, are classified as recurrent Novae.

During the last decade, several Novae (but only two symbiotic) have been detected by the Fermi-LAT (Large Area Telescope) at GeV $\gamma$-ray energies (Abdo et al.~2010, Ackermann et al.~2014, 
Franckowiak et al.~2018 and see also Table~S1 in 
Chomiuk et al.~2021). The spectra of these sources are usually well described by a simple power law
function, with the differential spectral index close to -2, and the exponential cut-off at a few to several GeV.
$\gamma$-ray emission appears with some delay after the initial optical flash. It lasts up to several 
to a few tens of days after explosion. In the case of the recurrent Nova RS Ophiuchi (RS Oph), exploded 
for the last time in August 2021, $\gamma$-ray emission is observed within the
GeV (Cheung et al.~2021) up to sub-TeV energy range (Acciari et al.~2022,  Aharonian et al.~2022).
The spectrum is well described by a broken power law function with the spectral index close to -2 in the GeV 
energy range. the spectrum steepens to about -4 at sub-TeV energies. On the other hand, $\gamma$-ray spectrum hardens with time after 
explosion. These results indicate that particles are accelerated in Novae at least up to TeV energies.
They likely contribute to the cosmic ray background in the Galaxy locally around the source (Acciari et al.~2022).  

$\gamma$-ray emission from Novae is expected to be appear either in the decay process of neutral pions produced by hadrons which collide with
the matter of the expending material in the nova shell, or by electrons which Inverse Compton up-scatter
soft radiation from the Nova photosphere. This radiation dominates at the early time after explosion 
(see e.g. Abdo et al. 2010, Sitarek \& Bednarek 2012, Martin \& Dubus 2013, Metzger et al. 2015, Ahnen et al. 2015, Vurm \& Metzger 2018, Martin et al. 2018).
The observation of the GeV $\gamma$-ray emission, at the early stage of the Nova explosion, suggests 
that this emission is likely produced by hadrons, when the shell is still dense. But the transfer of
energy from hadrons to radiation has to be rather efficient. On the other hand, the radiation field for relativistic
electrons is very dense,
allowing efficient transfer of energy from electrons to radiation. Other processes, e.g. the synchrotron
energy losses, can additionally extract energy from electrons, especially at the largest energies. 
This effect can be responsible for the change of the spectral index of the $\gamma$-ray emission from RS Oph in the
energy range between GeV emission and sub-TeV range. The leptonic model is consistent with the striking 
correlations
between the optical and the $\gamma$-ray light curves observed by Aydi et al.~(2020a) and Li et al.~(2017).
 
The Nova RS Oph belongs to the special class of recurrent symbiotic Novae. Already six explosions of this Nova 
have been observed with the recurrence period of the order of several years.   
The binary system RS Oph contains a massive WD (with the mass estimated on 1.35~M$_\odot$, i.e. close to 
the Chandrasekar limit, see Dobrzycka \& Kenyon~1994, Shore et al. 1996, Hachisu et al.~2009), 
and M0-2~III donor Red Giant (RG) with the mass 0.68-0.8~M$_\odot$  (Anupama \& Miko\l ajewska~1999) 
and the luminosity $\sim 10^3$ times larger than the luminosity of the Sun. 
The main part of the RG surface has the temperature of 3600~K (Pavlenko et al. 2021). Its radius is estimated on
$67^{+19}_{-16}$~R$_\odot$, where $R_\odot$ is the radius of the Sun (Dumm \& Schild~1998). The axis of the
binary system is inclined at the angle  
39$^{+1}_{-10}$ degrees to the observer's line of sight (Ribeiro et al. 2009). The mass transfer 
to the WD is estimated in the range $3.7\times 10^{-8}$~M$_\odot$~yr$^{-1}$ (Schaefer~2009) 
and $10^{-6}$~M$_\odot$~yr$^{-1}$ (Iijima~2006). Then, the upper limit on the mass expelled in the Nova shell
is scaled by the recurrence time of Nova RS Oph, i.e it is in the range $5.6\times 10^{-7}$~M$_\odot$
to $1.5\times 10^{-5}$~M$_\odot$. 
The RG is expected to produce the equatorial wind with the mass loss rate $\sim 10^{-7}$~M$_\odot$~yr$^{-1}$
and the velocity 40~km~s$^{-1}$ (Wallerstein 1958). The kinetic and geometrical structure of 
the Nova ejecta looks very complicated. 
The velocity of ejecta, at about one day after explosion in 2006, was measured on $\sim 4000 - 7500$~km~s$^{-1}$ 
(Buil et al.~2006).
The composite structure of the ejecta from the Nova RS Oph, at a few hundred days after explosion in 2006
as observed by the  Hubble Space Telescope, shows two component symmetric flow, with the low velocity high density 
equatorial region and the high velocity polar region. A true expansion velocity for the material at 
the poles was measured on $5600\pm 1100$~km~s$^{-1}$ (Bode et al.~2007). 

Based on these observations, it is concluded that the ejected shell of material moves with different velocities
in the equatorial and the polar regions of the binary system. In the simplest two-component scenario, 
the Nova shell is composed of the equatorial slow and dense region and the fast, freely expending polar region  
(see Shore et al.~2011, 2013 and Mason et al. 2018). Moreover, there are evidences of multiple ejections. 
Initial slow and dense material is reached by the fast wind
from the WD surface formed at some time after initial explosion (Aydi et al.~2020b).

Here we investigate the consequences of the hypothesis that electrons are continuously accelerated
in the expending Nova shell applying a simple two-component geometrical model.
It is assumed that the Nova exploded initially spherically symmetric. But due to the axially symmetric wind of 
the Red Giant,
the shell is significantly decelerated in the equatorial region of the binary system but propagates almost
freely in the polar regions. The presence of non-thermal particles, and the general two-component structure
of the outflow, is supported by the observations of the non-thermal radio emission on a time scale of a month
after explosion (e.g. Eyres et al.~2009). Such radio emission can  last for years, as observed in the case 
of Nova V445 Pup (Nyamai et. al.~2021), indicating that particles can be accelerated in the Nova shells 
for a similar time scale.

We calculate the $\gamma$-ray spectra produced in the Inverse Compton (IC) process by electrons
accelerated continuously during the propagation of the Nova shell. The equilibrium spectra of electrons are
calculated at an arbitrary moment after the initial explosion taking into account different energy loss processes
of electrons. Due to weaker energy losses late after explosion, electrons can reach multi-TeV 
energies, producing TeV $\gamma$-rays mainly in the comptonization process of the RG radiation.
In order to test the hypothesis on the extended acceleration of electrons in the expending shell of Nova RS Oph, 
we compare predicted $\gamma$-ray fluxes with the sensitivities of the present 
and future Cherenkov telescopes.

\section{Structures of nebulae around recurrent Novae}

Recurrent Novae are a sub-class of Novae which show regular explosions with the recurrence period shorter than 
about hundred years. Up to now, the largest number of explosions (known 6) have been observed in the case of 
the Nova RS Oph. RS Oph is a specially  interesting Nova since it belongs to a class of 
recurrent symbiotic Novae in which the White Dwarf forms a binary system with a RG star. 
In such Novae, the geometry of explosion is particularly complex. The shell of expelled matter 
(even if ejected isotropically) can be modified due to the
interaction with the RG wind which is concentrated in the orbital plane of the binary system. 
Therefore, a part of the Nova shell can be significantly decelerated in the orbital plane of the binary system. 
On the other hand, in polar regions, the shell expends with a constant velocity. In fact, such two-component 
expansion of the Nova shell has been observed in the case of the previous explosion of RS Oph in 2006 
(see e.g. Ribeiro et al. 2009). Such geometrical two component model (originally proposed by Shore et al.~2011,
2013 and Mason et al. 2018) is considered in our work.
Another complication might be due to multiple ejection of material by the Nova with 
different velocities (Aydi et al.~2020b). The material of initial ejection is expected to move with the lower 
velocity than the latter formed wind from the WD. 
As a result of their collisions, the shell is additionally energized.

In the case of a recurrent Nova RS Oph, we consider a simple two component geometrical structure for the Nova 
shell. We accept that the material is ejected isotropically from the White Dwarf surface with the mass 
of the order of $M_{\rm sh} = 10^{-6}\dot{M}_{-6}$~M$_\odot$ and with the velocity of the order of 
$v_{\rm sh} = 3000v_8$~km~s$^{-1}$. 
The total kinetic energy of the expending shell of Nova  is then 
$L_{\rm kin} = 0.5M_{\rm sh}v_{\rm sh}^2 = 9\times 10^{43}\dot{M}_{-6}v_8^2$~erg. During the recurrence time scale, 
this shell is expected to move with the constant velocity in the polar region of the binary system but 
it can be significantly decelerated in the equatorial region of the RS Oph binary system on a time scale of months. 
The deceleration is due to the entrainment of the Red Giant wind by the shell.
The Red Giant wind is assumed to be expelled within the equatorial region of the binary system, i.e.
within a part of the whole sphere equal to $\Omega_{\rm RG}$. 
The Red Giant wind is expelled with the rate $\dot{M}_{\rm RG} = 10^{-7}M_{-7}~M_{\odot}$~yr$^{-1}$, 
and the velocity $v_{\rm RG} = 10^6v_6$~cm~s$^{-1}$.

After ejection of the material by the WD, the nuclear energy is still generated in 
the remnant layer of matter on the WD surface. This energy is irradiated from the optically thick photosphere 
with the initial luminosity over the Eddington luminosity of the Solar mass object.
During the time of several to a few tens of days, the photosphere becomes transparent to the soft thermal 
emission from the WD surface. This emission appears in the soft X-rays since the typical temperature of 
the surface of the WD is of the order of
a few $10^5$~K. During the initial optically thick stage of the layer of matter on the WD surface, also 
a strong wind is expelled from the WD. This wind  collides with the Nova shell, 
additionally energizing the material in the shell, already at some time after initial Nova explosion. 
We are interested in the processes occurring within the shell from
the moment of this additional energization by the fast wind from the WD.  

Recent observations of the sub-TeV $\gamma$-ray emission from RS Oph (Acciari et al.~2022, Aharonian et al.~2022)
indicate that particles (electrons, hadrons) have to be accelerated during the Nova explosion. 
The site and mechanism of their acceleration is not presently known. Particles might be accelerated either in 
the wind or in the collision region between the Nova wind 
with the ejecta or in the expending Nova shell. We consider a scenario in which electrons are accelerated 
continuously in the Nova shell. At early phase after ejection, lasting from several days to a few weeks,
accelerated electrons cool mainly on the IC process by scattering radiation from the nearby photosphere and 
from the RG. At latter phase, electrons lose a part of the energy mainly of their synchrotron process. 
Electrons  are captured in the shell by the magnetic field being advected from the central part of the Nova binary
system with 
the expending shell. If the shell moves with the constant velocity, relativistic electrons are transported
to the distance from the WD equal to 
$R_{\rm sh} = \tau_{\rm rec}v_{\rm sh}\approx  1.4\times 10^{17}\tau_{15}v_8$~cm, where the recurrence time 
of the Nova is $\tau_{\rm rec} = 15\tau_{15}$~years. 
Electrons are still immersed in the radiation of the RG star producing $\gamma$-rays in the Inverse Compton process.
We expect that this IC $\gamma$-ray component should appear at larger energies than sub-TeV $\gamma$-ray 
emission observed a few days after explosion, since electrons are expected to reach 
larger energies due to longer time scale for their energy losses on the synchrotron process and 
also due to the larger dynamical time scale of the shell.

\begin{figure}
\vskip 6.truecm
\includegraphics{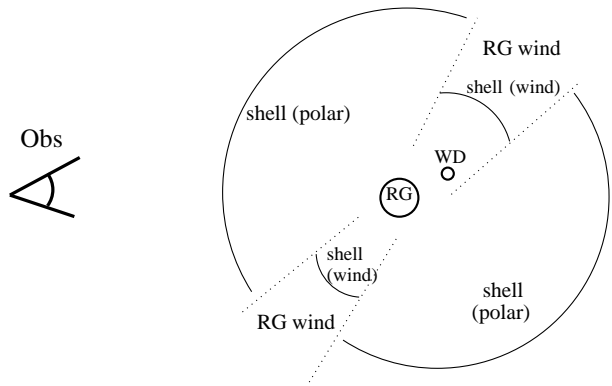}
\caption{Schematic representation of the symbiotic Nova RS Oph. The explosion on the White Dwarf surface 
occurs spherically symmetric. After some time, a part of the Nova shell is efficiently decelerated 
in the equatorial region of the binary system due to its interaction with the Red Giant wind. 
The shell in the polar region expends freely.  The Observer sees the equatorial part of the two-component shell
at the angle $\sim$40$^\circ$. Electrons are accelerated in both regions to different maximum energies
(due to different magnetic field strengths).  These two components of the shell are also at other distances 
from the soft radiation of 
the Red Giant. Electrons are accelerated isotropically with a constant rate in time in both regions.}
\label{fig1}
\end{figure}
\section{Acceleration of electrons in the Nova shell}

We follow the standard scenario for the explosion of Nova. Ejected material forms a shell
with some thickness $\Delta R =\beta R$, where $R$ is the shell radius and 
the scaling factor is $\beta = 0.1\beta_{-1}$. 
Particles are expected to be accelerated in the region of the shell.
How this process occurs in the case of Nova explosions is at present an open question.
In principle, two sites can be investigated, i.e. regions between the ejecta and the surrounding medium
or between the shell material and the Nova wind. We concentrate on the first possibility since it provides
eventual mechanism for acceleration of particles on a time scale of the propagation of the shell.
The theory of shock acceleration argues that relativistic particles can take
as much as $\sim 10-20\%$ of the kinetic energy of the shock. The acceleration process of particles 
can be limited either by their energy losses or the dynamical time scale of the shock.
Therefore, in order to estimate the maximum energies to which electrons are accelerated, we consider a simple 
model for the magnetic field structure within the expending shell. This magnetic field determines the main
processes for the energy gain and energy losses of particles in the shell.  

We estimate the strength of the magnetic field in the expending shell of Nova assuming some level of 
equipartition between the kinetic energy density of the ejected shell and the energy density of 
the magnetic field. 
The radius of the WD in the Nova RS Oph is $R_{\rm WD} = 2.3\times 10^8$~cm (see formula in Nauenberg 1972, 
for the WD mass 
$\sim 1.35$~M$_\odot$, see Dobrzycka \& Kenyon~1994, Shore et al. 1996, Hachisu, Kato, Luna~2009).
The distance from the WD is expressed in units of the stellar radius according to 
$R = r R_{\rm WD}$. The density of the matter in the ejected shell is
\begin{eqnarray}
n_{\rm sh} = {{M_{\rm sh}}\over{4\pi \beta R^3m_{\rm p}}}\approx {{8.2\times 10^{25} M_{-6}}\over{\beta_{-1}r^3}}
\approx {{5.8\times 10^{10} M_{-6}}\over{\beta_{-1}t_{\rm d}^3v_8^3}}~~~{{\rm cm}^{-3}},
\label{eq1}
\end{eqnarray}
\noindent
where $r = v_{\rm sh}t/R_{\rm WD}\approx 1.12\times 10^5t_{\rm d}v_8$, the time $t = 8.6\times 10^4t_{\rm d}$ s,
and $m_{\rm p}$ is the proton mass.
The magnetic field strength, at some level of equipartition with the kinetic energy density of the shell, 
is obtained from $\alpha n_{\rm sh}m_{\rm p}v_{\rm sh}^2/2 = B^2/(8\pi)$,
where $\alpha = 10^{-2}\alpha_{-2}$ is so called equipartition coefficient. The magnetic field is
\begin{eqnarray}
B = (4\pi \alpha n_{\rm sh} m_{\rm p})^{1/2}v_{\rm sh}\approx 
32v_8({{\alpha_{-2}M_{-6}}\over{\beta_{-1}t_{\rm d}^3v_8^3}})^{1/2}~~~{\rm G}.
\label{eq2}
\end{eqnarray}
\noindent
For the scaling value of the parameter $\alpha \sim 0.01$, this prescription is generally consistent with 
the estimates of the magnetic field in the shell during previous explosion
of RS Oph obtained at 20 days after explosion from the equipartition arguments 
based on the radio observations, i.e. $B\sim (0.08 - 0.11)$ G (Rupen et al.~2008).

On the other hand, electrons are accelerated at the energy gain rate which can be parametrised by 
\begin{eqnarray}
\dot{E}_{\rm acc} = {{\xi cE}\over{R_{\rm L}}}\approx 0.1\xi_{-5}B
\approx 3.2\xi_{-5}v_8({{\alpha_{-2}M_{-6}}\over{\beta_{-1}t_{\rm d}^3v_8^3}})^{1/2}~~~{\rm {{GeV}\over{s}}},
\label{eq3}
\end{eqnarray}
where $\xi = 10^{-5}\xi_{-5}$ is the acceleration coefficient which is estimated on 
$\xi\sim (v_{\rm turb}/c)^2\sim  10^{-5}v_8^2$, where $v_{\rm turb}\sim 0.3v_{\rm sh}$ is the turbulent 
velocity of scattering centres above the shock of the order of the fluid flow for the strong shock.
We use this prescription for the slow acceleration of electrons which is characteristic for the second
order Fermi acceleration process. However, in Sect. 6, we discuss the results for the prescription 
for the fast acceleration 
of electrons, characteristic for the first order Fermi acceleration process. In this case, $\gamma$-ray spectra
only slightly extend to larger energies since at 2 years after explosion the dynamical time scale of the shell
becomes shorter than the synchrotron cooling time scale of electrons.

\begin{figure}
\vskip 10.8truecm
\includegraphics{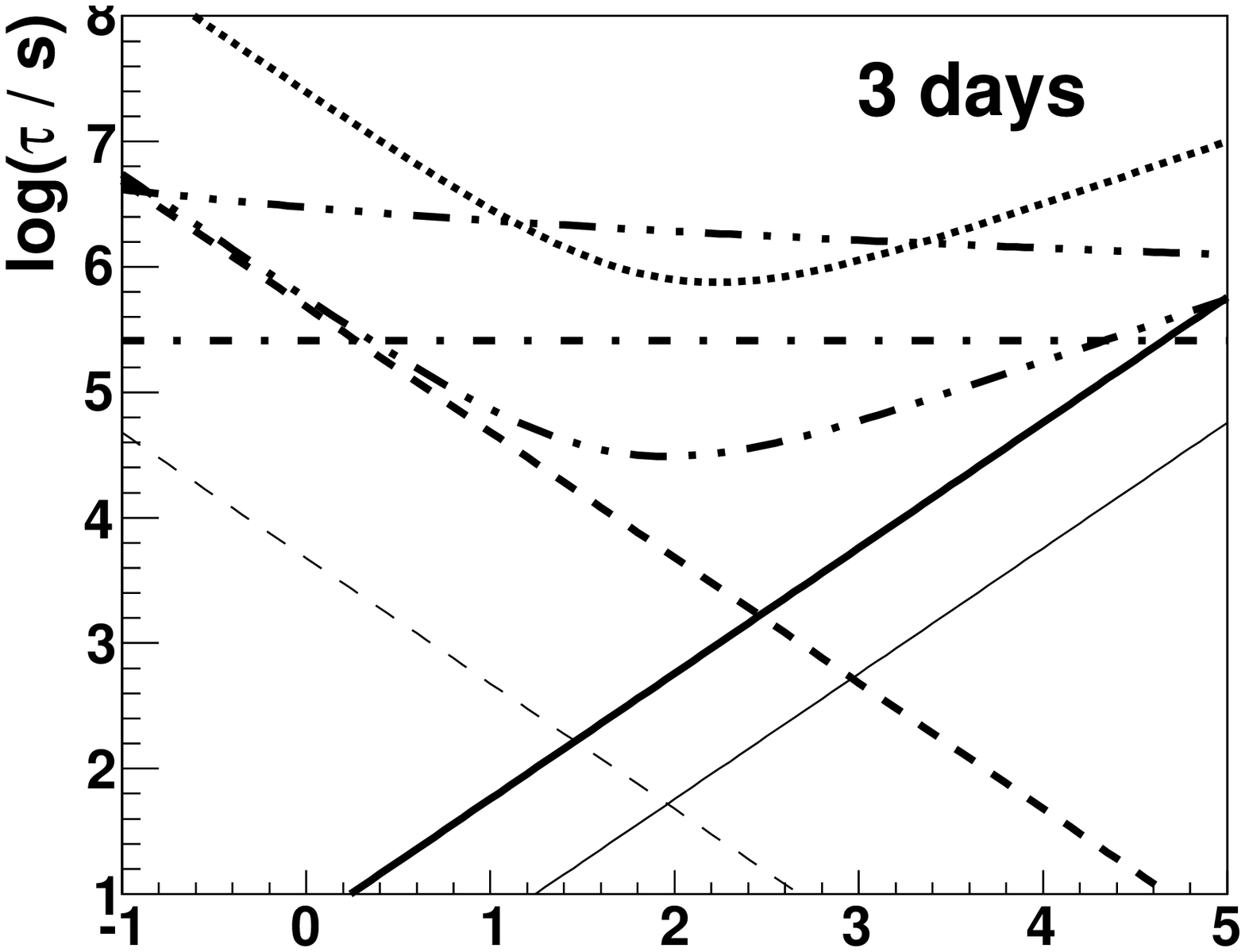}
\includegraphics{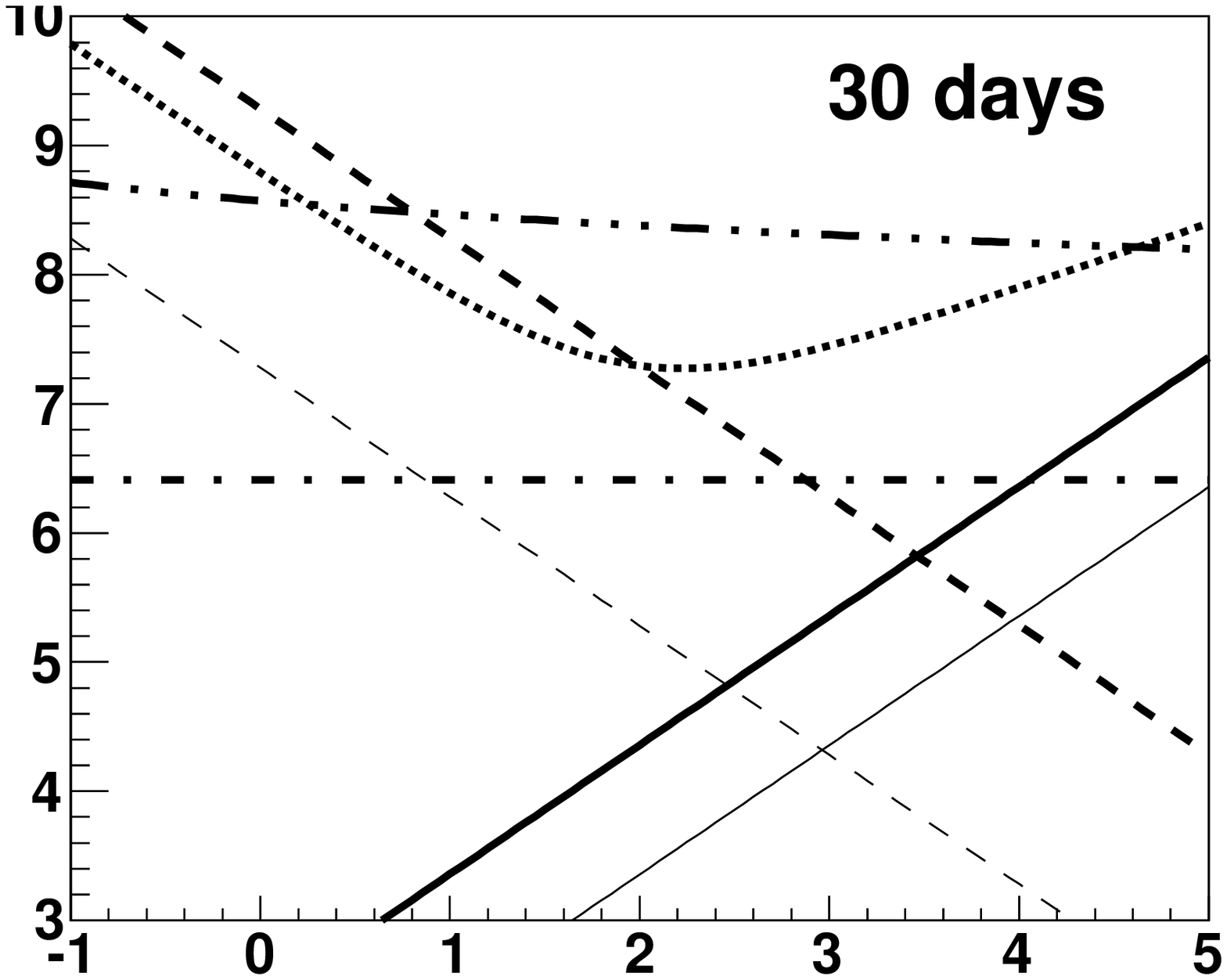}
\includegraphics{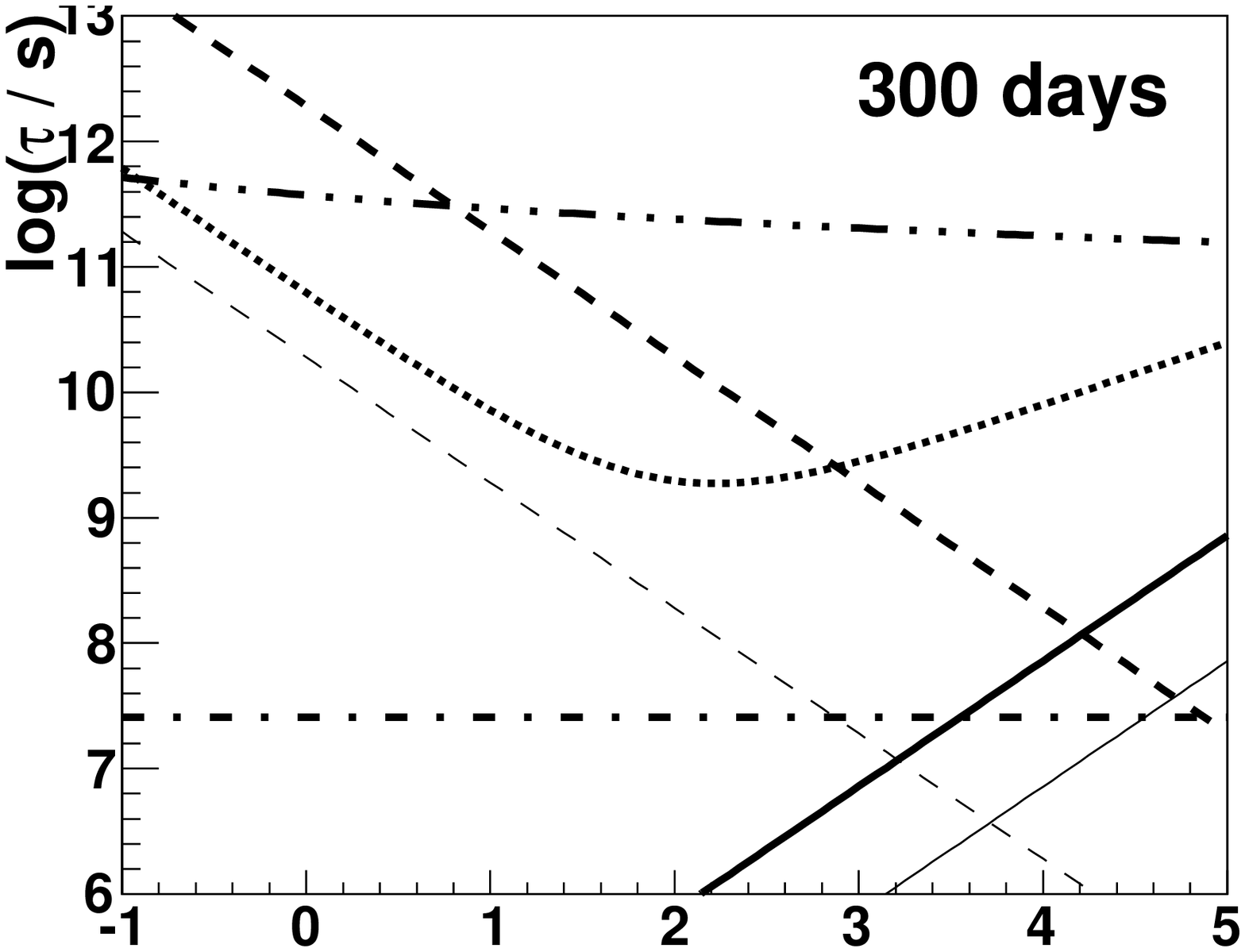}
\includegraphics{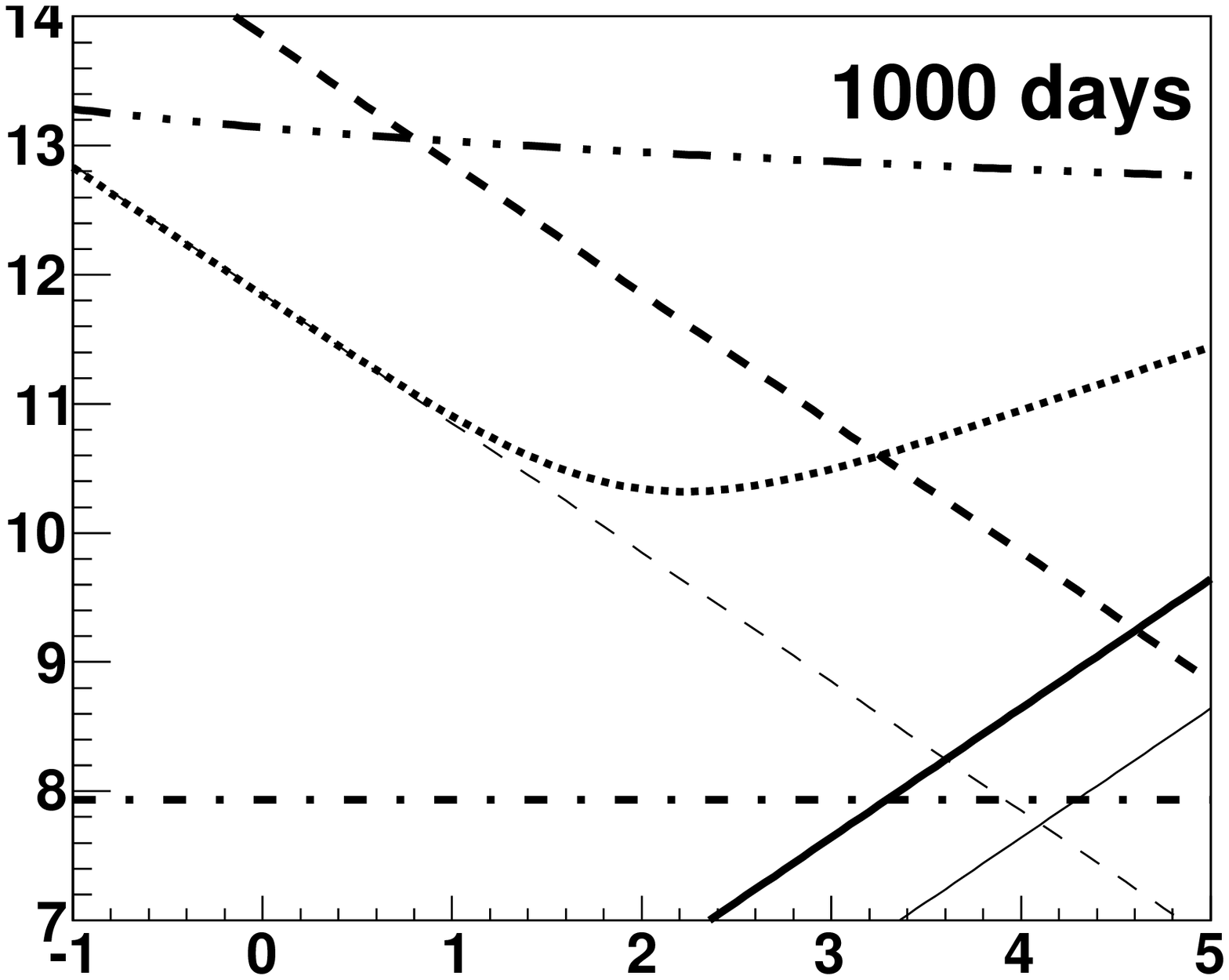}
\includegraphics{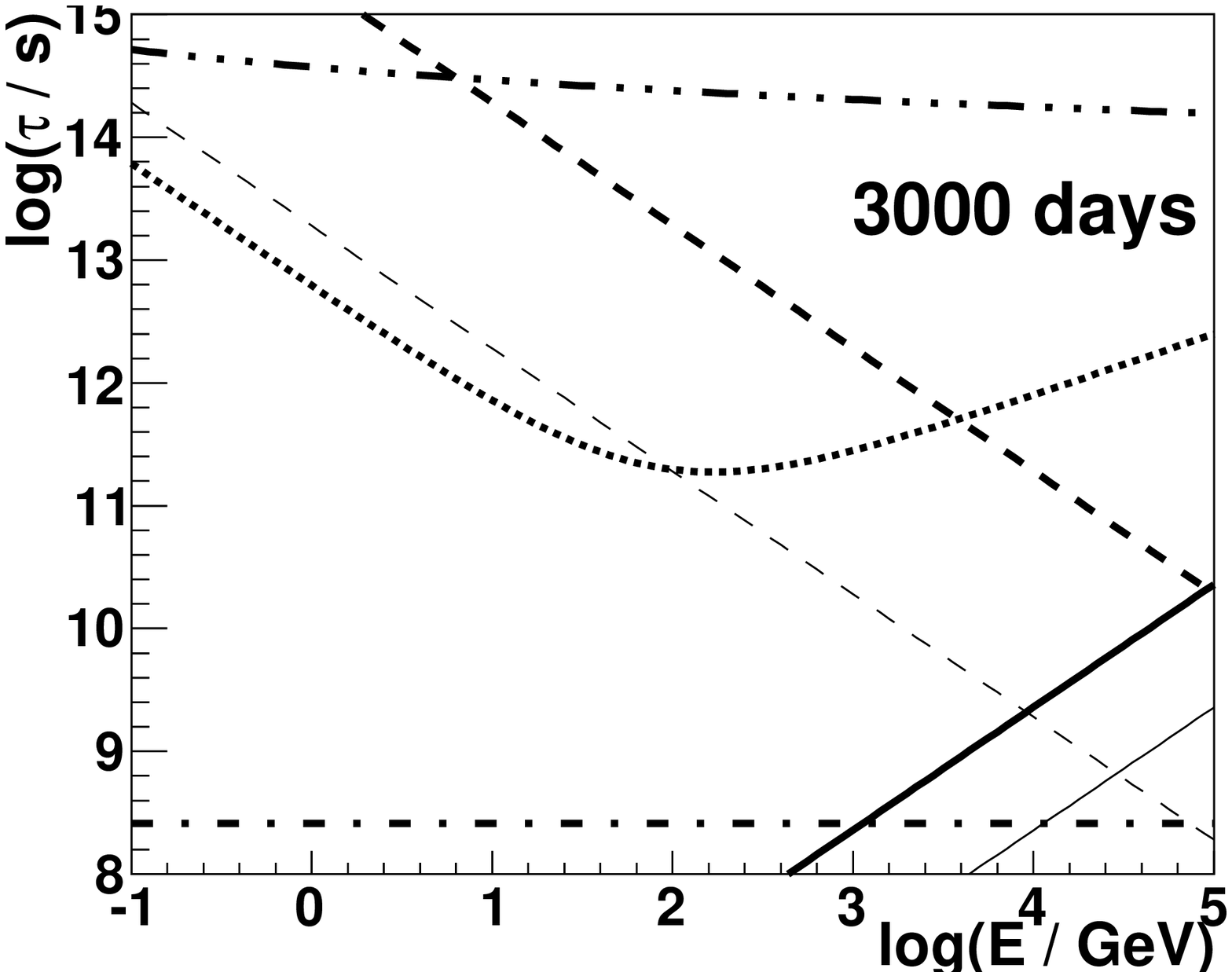}
\includegraphics{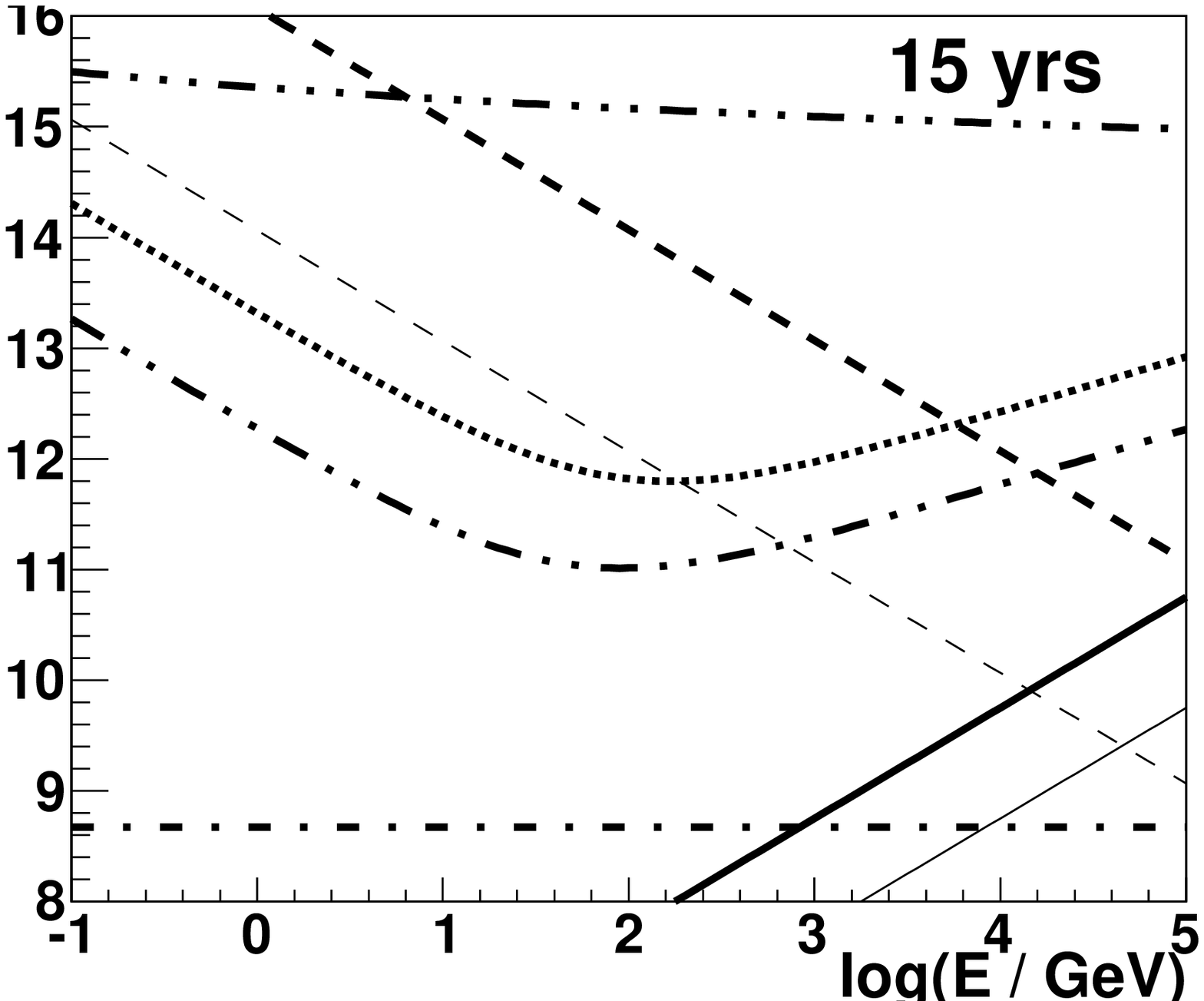}
\caption{The time scales, $\tau$, are shown for different energy gains and loss processes of relativistic electrons  
in the shell of the recurrent Nova RS Oph at different moments after explosion (equal to 3 days, 30 days, 
300 days, 1000 days, 3000 days, and 15 years). The following time scales are marked by: 
the acceleration time scale for the equipartition coefficient $\alpha = 0.01$ 
(thin solid curve) and $10^{-4}$ (thick solid), the time scale for the synchrotron energy losses of electrons 
are marked by dashed lines, for $\alpha = 0.01$ (thin) and for $10^{-4}$ (thick), the dynamical time scale 
(dot-dashed), the bremsstrahlung time scale (dot-dot-dot-dashed), the IC time scale in the 
Red Giant radiation field (dotted) and in the radiation field from the Nova photosphere as observed at 3 days 
and 15 years after explosion of the Nova (dot-dot-dashed). The parameters of the Nova photosphere and the Red Giant are 
reported in the main text.}
\label{fig2}
\end{figure}
\begin{figure}
\vskip 4.8truecm 
\includegraphics{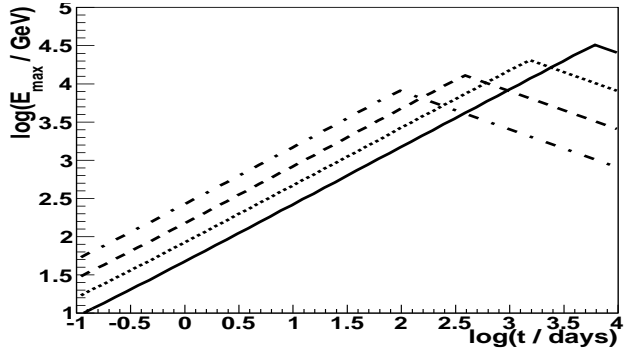}
\caption{Characteristic energies of electrons accelerated in  the freely expanding shell of Nova 
as a function of time after initial explosion, for a few different strengths of the magnetic field
in the shell defined by the equipartition coefficient $\alpha = 0.1$ (solid curve),
$10^{-2}$ (dotted), $10^{-3}$ (dashed), and $10^{-4}$ (dot-dashed). The mass of the Nova shell is assumed 
to be $M_{\rm sh} = 10^{-6}$~M$_\odot$, the shell velocity is $v_{\rm sh} = 6\times 10^8$~cm~s$^{-1}$,
and the thickness of the shell is $\beta = 0.1$.}
\label{fig3}
\end{figure}

The energy loss rate of electrons on the synchrotron process depends on the energy density of the magnetic field,
\begin{eqnarray}
\rho_{\rm syn} = {{B^2}\over{8\pi}}\approx 2.6\times 10^4v_8^2(\alpha_{-2}M_{-6}/\beta_{-1}t_{\rm d}^3v_8^3)
~~~{\rm {{GeV}\over{cm^{3}}}}.
\label{eq4}
\end{eqnarray}
\noindent
It is
\begin{eqnarray}
\dot{E}_{\rm syn} = {{4}\over{3}}c\rho_{\rm syn} \sigma_{\rm T}\gamma^2 \approx 
1.4\times 10^{-3}v_8^2{{\alpha_{-2}M_{-6}}\over{\beta_{-1}t_{\rm d}^3}} E_{\rm GeV}^2~~~{\rm {{GeV}\over{s}}}.
\label{eq5}
\end{eqnarray}
\noindent
By comparing the energy gain rate with the synchrotron energy loss rate, we obtain the maximum 
energies of electrons
\begin{eqnarray}
E_{\rm max}^{\rm syn} = 34(\xi_{-5}/v_8)^{1/2}({{\beta_{-1}t_{\rm d}^3}\over{\alpha_{-2}v_8^3M_{-6}}})^{1/4}
~~~{\rm GeV}.
\label{eq6}
\end{eqnarray}
Close to the central engine, the above process determines the maximum energies of electrons. 
By comparing the maximum energies of accelerated electrons with the maximum energy of the $\gamma$-ray
photons equal to $\sim$500~GeV (Aharonian et al.~2022), i.e. 
$E_{\rm max}^{\rm syn} > E_\gamma^{\rm max}\approx 500$~GeV, we constrain the equipartition 
coefficient $\alpha < 2\times 10^{-5}\beta_{-1}(\xi_{-5}v_8)^2t_{\rm d}^3v_8^3/M_{-6}$.
For larger magnetization, the synchrotron energy losses do not allow acceleration of electrons
to the observed $\gamma$-ray energies. Therefore, shells cannot be strongly magnetized already a few days
after formation. However, note that this constraint strongly depends on the time 
after explosion.

At large distances from the WD, the energy losses of electrons becomes too low (due to the strong drop of 
the magnetic 
field during the expansion of the Nova shell) to balance the energy gains from the acceleration mechanism.
Then, the maximum energies of electrons are limited by the dynamic time scale of the Nova shell equal to
$\tau_{\rm dyn} = R_{\rm sh}/v_{\rm sh} = 8.6\times 10^4t_{\rm d}$~s. Then, the electron energies  are obtained
from the comparison of the acceleration time scale, $\tau_{\rm acc} = E_{\rm e}/\dot{E}_{\rm acc}$, with
$\tau_{\rm dyn}$, 
\begin{eqnarray}
E_{\rm max}^{\rm dyn}\approx 2.75\times 10^5\xi_{-5}v_8({{\alpha_{-2}M_{-6}}
\over{\beta_{-1}t_{\rm d}v_8}})^{1/2}~~~{\rm GeV}.
\label{eq7}
\end{eqnarray}
\noindent
The time scales for different energy loss processes  of electrons in the Nova shell, the dynamical time scale of 
the shell and the acceleration time scale of electrons in the shell, are shown in Fig.~2, 
for specific time after the Nova explosion.
For the considered range of the magnetic field strengths in the shell, the synchrotron energy loss process 
dominates
for the most energetic electrons accelerated in the Nova shell. However, at early time after explosion, the IC 
scattering of soft radiation from the Nova photosphere becomes dominant mechanism for the electron 
energy losses.
At the later time after explosion, the energy losses on the Inverse Compton scattering of soft radiation from 
the RG start to dominate, accept the moment of the next explosion at 15 years after original explosion. 
At the late stage of the Nova, the
electrons with energies below  $\sim$TeV, lose energy mainly on the $\gamma$-ray production in the IC process.  
As we noted above, the maximum energies of electrons are determined by the balance between the energy gains 
from the 
acceleration mechanism and energy losses on the synchrotron process or on the dynamical time scale. 
In Fig.~3, we show how those maximum energies of electrons evolve with time after initial 
explosion of the Nova, for fixed parameters of the Nova shell and different values of the magnetic 
field strength which is described by the 
equipartition coefficient $\alpha$. Electrons reach the maximum energies at the intermediate time after explosion. 
The maximum energies of electrons at the early time are constrained by the synchrotron energy losses and, 
at late time, by  the dynamical time of the Nova shell.

\section{Equilibrium spectrum of electrons within the shell}

We assume that electrons are injected into the Nova shell from some acceleration mechanism. The acceleration process starts just after the Nova explosion and continue, at a constant rate in time, up to the observation time '$t_{\rm obs}$'. Electrons are injected with the power law 
spectrum to characteristic maximum energy $E_{\rm max}$ (estimated by Eq.~6 or Eq.~7). It is assumed that electrons are confined 
in the shell by the magnetic field. The equilibrium spectrum of electrons within the shell at the time '$t_{\rm obs}$' is obtained by solving the continuity equation for electrons of the type discussed by Blumenthal \& Gould (1970, see Eq. 5.2).  However, since the spectrum of electrons evolve
within the shell in which physical conditions change in a complicated way in time
we solve this transport equation for electrons numerically.
We create a grid in the energy of injected electrons (in fact in log $E_e$) and the time.
We do not consider any additional acceleration of electrons after initial injection at the time 't'.
The contribution to the electron spectrum at the time '$t_{\rm obs}$' from freshly injected electrons at the time $t < t_{\rm obs}$
is calculated numerically by subtracting their energy losses on specific radiation
processes (synchrotron, bremsstrahlung and inverse Compton). These different energy losses are described in Sect. 4 and below.
In the case of the average energy losses of electrons on the inverse Compton process, we use the approximate formula which is also valid in the Klein-Nishina regime (Moderski et al.~2005).
The numerical method is used due to complicated dependence of the energy losses of electrons on time. These losses are 
determined by varying physical parameters during propagation of the shell (radiation field, magnetic field, density of matter).
However, we neglect the energy losses of electrons on the adiabatic expansion of the shell since this process is difficult to
consider on the present stage of knowledge. In fact, the shell might expand in time but also contract under the external
pressure of the cosmic space or the wind from the White Dwarf. 
Therefore, it is considered that only radiation processes will determine the equilibrium spectrum of electrons at a specific propagation time of the shell. We determine the effect of radiation processes on the electron's spectrum at a specific distance
$R_{\rm sh}$ by calculating the energy loss time scales, 
$\tau_{\rm syn, IC, br} = E_{\rm e}/\dot{E}_{\rm syn, IC, br}$. Those time scales are dependent on the 
energy density of the magnetic field, different types of soft radiation and the matter in the shell.

The energy density of the soft radiation field from the RG is     
\begin{eqnarray}
\rho_{\rm RG} = aT_{\rm RG}^4\approx {{7.2\times 10^{11}T_3^4R_{13}^2}\over{r^2}}\approx 
{{57T_3^4R_{13}^2}\over{(v_8t_{\rm d})^2}}
~~~{\rm {{GeV}\over{cm^{3}}}},
\label{eq8}
\end{eqnarray}
valid for $r > R_{\rm RG}/R_{\rm WD}$. The radius of the RG is scaled by $R_{\rm RG} = 10^{13}R_{13}$ cm, 
its surface temperature is $T_{\rm RG} = 3\times 10^3T_3$ K, and $a$ is the radiation constant.

On the other hand, the soft radiation from the Nova photosphere dominates at the early time after the Nova 
explosion, but
also at the recurrence period of the Nova explosion (see Fig.~2). The energy density of radiation from 
the Nova photosphere is given by
\begin{eqnarray}
\rho_{\rm ph} = aT_{\rm ph}^4\approx {{8.9\times 10^{13}T_4^4R_{13}^2}\over{r^2}}\approx 
{{7.1\times 10^{3}T_4^4R_{13}^2}\over{(v_8t_{\rm d})^2}}~~~{\rm {{GeV}\over{cm^{3}}}},
\label{eq9}
\end{eqnarray}
\noindent
where the temperature of the photosphere is $T_{\rm ph} = 10^4T_4$~K, and its radius 
is $R_{\rm ph} = 10^{13}R_{13}$~cm.
The radiation field from the Nova photosphere dominates at several days after explosion.
The evolution of the optical emission from the Nova RS Oph, as observed during the explosion in August 2021 
(see Fig.~1 in Acciari et al.~2022), can be 
approximated by $\log (L/L_\odot) = 4.90 - 0.23t_{\rm d}$ in the period between 0-4 days and 
$\log (L/L_\odot) = 3.98 - 0.056(t_{\rm d} - 4)$ in the period between 4-15 days, where $t_{\rm d}$ is the time 
from explosion in days. We accept that the radiation comes from the Nova photosphere with the temperature
$T_{\rm ph} = 5500$~K. It does not change significantly in time. Then, the photo-spheric radius can be 
estimated from $R_{\rm phot} = (L/4\pi \sigma_{\rm SB}T_{\rm phot}^4)^{1/2}$, where $\sigma_{\rm SB}$ 
is the Stefan-Boltzmann constant. The radiation field, at the specific distance 
of the Nova shell, is assumed to be of the black body type diluted by a factor of $(R_{\rm phot}/R_{\rm sh})^2$.

The cooling time scales of electrons on the IC process (in the general case) in those different radiation fields 
are calculated from the approximate formula given in Sect.~2.1 in Moderski et al.~(2005).
We also calculate the energy losses of electrons on the bremsstrahlung process in a completely ionized hydrogen using the formula, 
\begin{eqnarray}
\dot{\rm E}_{\rm brem}\approx {{c n_{\rm sh}m_{\rm p}E}\over{X_{\rm 0}}} F\approx 
{{4.5\times 10^{-5} M_{-6}E_{\rm GeV}}\over{\beta_{-1}v_8^3t_{\rm d}^3}}F~~~{\rm {{GeV}\over{s}}},
\label{eq10}
\end{eqnarray}
where
$X_{\rm 0} = 62.8$~g~cm$^{-2}$ is the radiation length in a neutral Hydrogen. The correction factor, 
$F = \ln{(2E/m_ec^2)}/\ln{183}$, is introduced for the bremsstrahlung energy losses in the case of ionized hydrogen. 
The detailed formula for this process can be found in Haug~(2004).
Note that the bremsstrahlung energy losses are usually clearly lower in respect to other energy loss processes
(see Fig.~2).
Then, the cooling time scale on the bremsstrahlung  process can be approximated by
\begin{eqnarray}
\tau_{\rm brem} = E/\dot{\rm E}_{\rm brem}\approx 2.2\times 10^4 \beta_{-1}v_8^3t_{\rm d}^3/(M_{-6}F)~~~s.
\label{eq11}
\end{eqnarray}
\noindent
The energy loss time scales of electrons on different processes, their energy gains 
from the acceleration mechanism and the dynamical time scale at different distance of the shell from the WD
are shown in Fig.~2.

Electrons, confined within the shell, 
lose energy on different energy loss processes. Since the conditions within the shell change significantly
in time (and propagation distance),
we develop a numerical code in order to calculate the equilibrium spectrum of electrons at a specific distance
(and propagation time) from the WD. 
At first, we consider the simple scenario in which the shell is expanding freely, without any 
deceleration by the medium surrounding the jet or any energy gain from the wind of the White Dwarf.

\begin{figure*}
\vskip 4.5truecm
\includegraphics{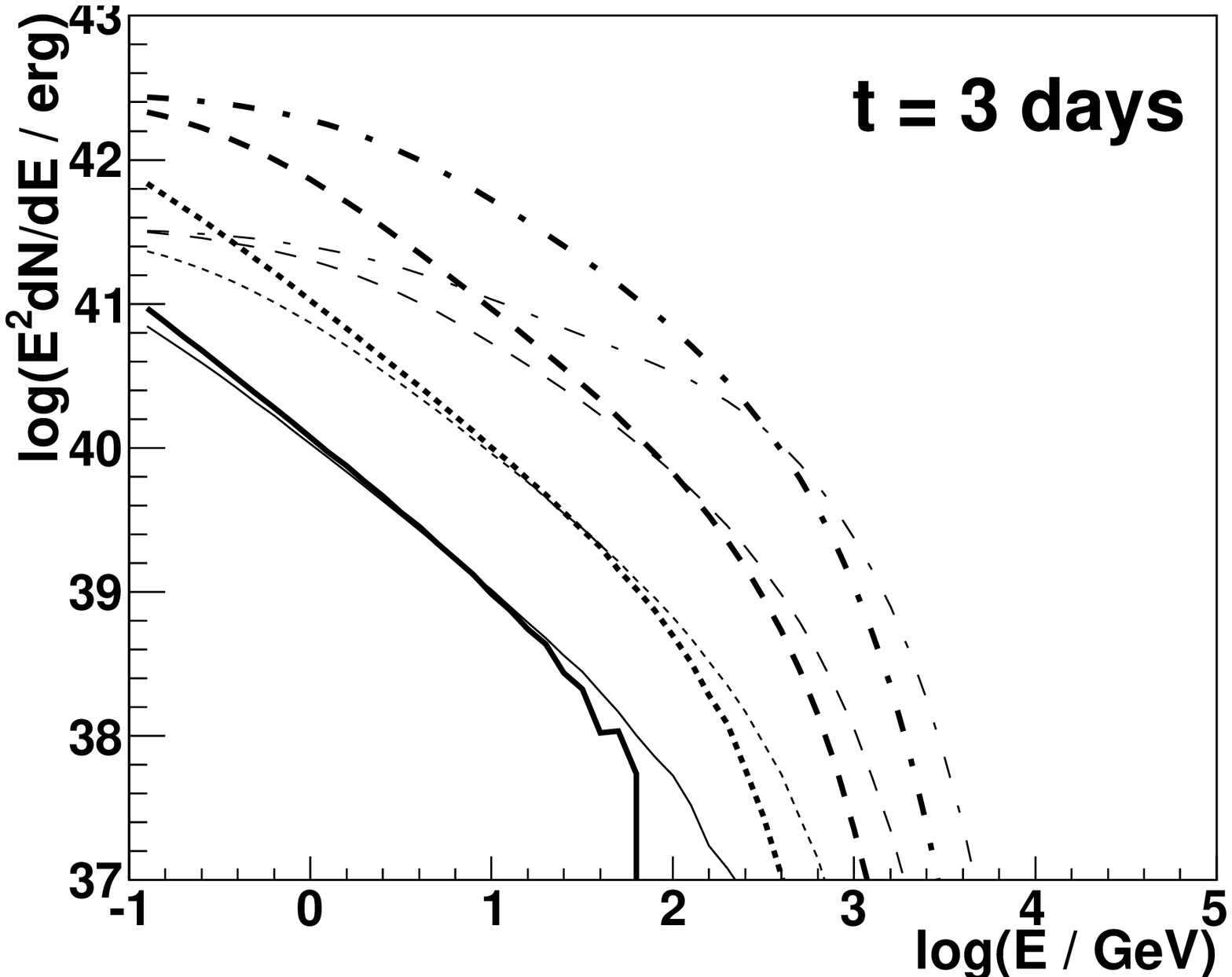}
\includegraphics{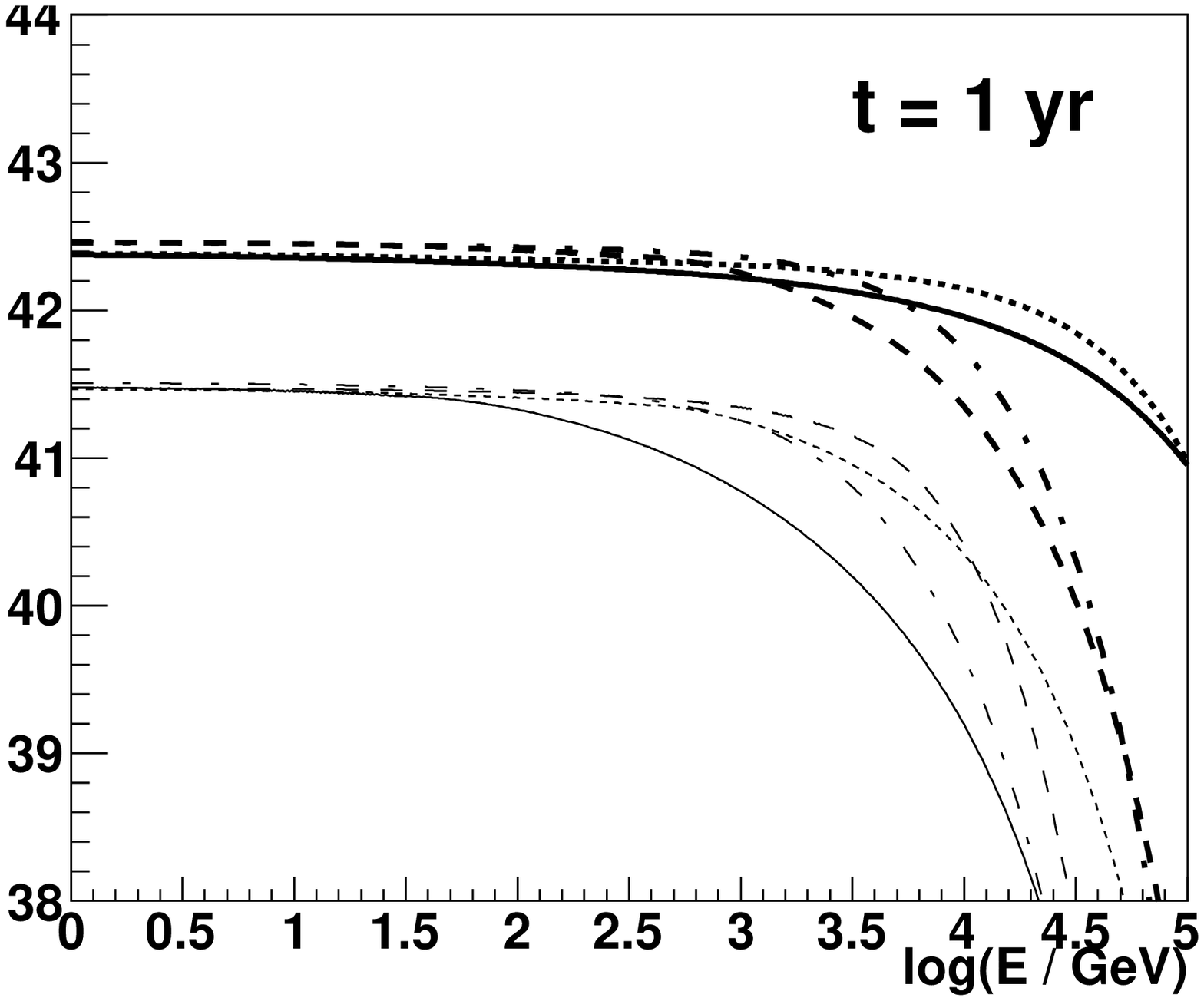}
\includegraphics{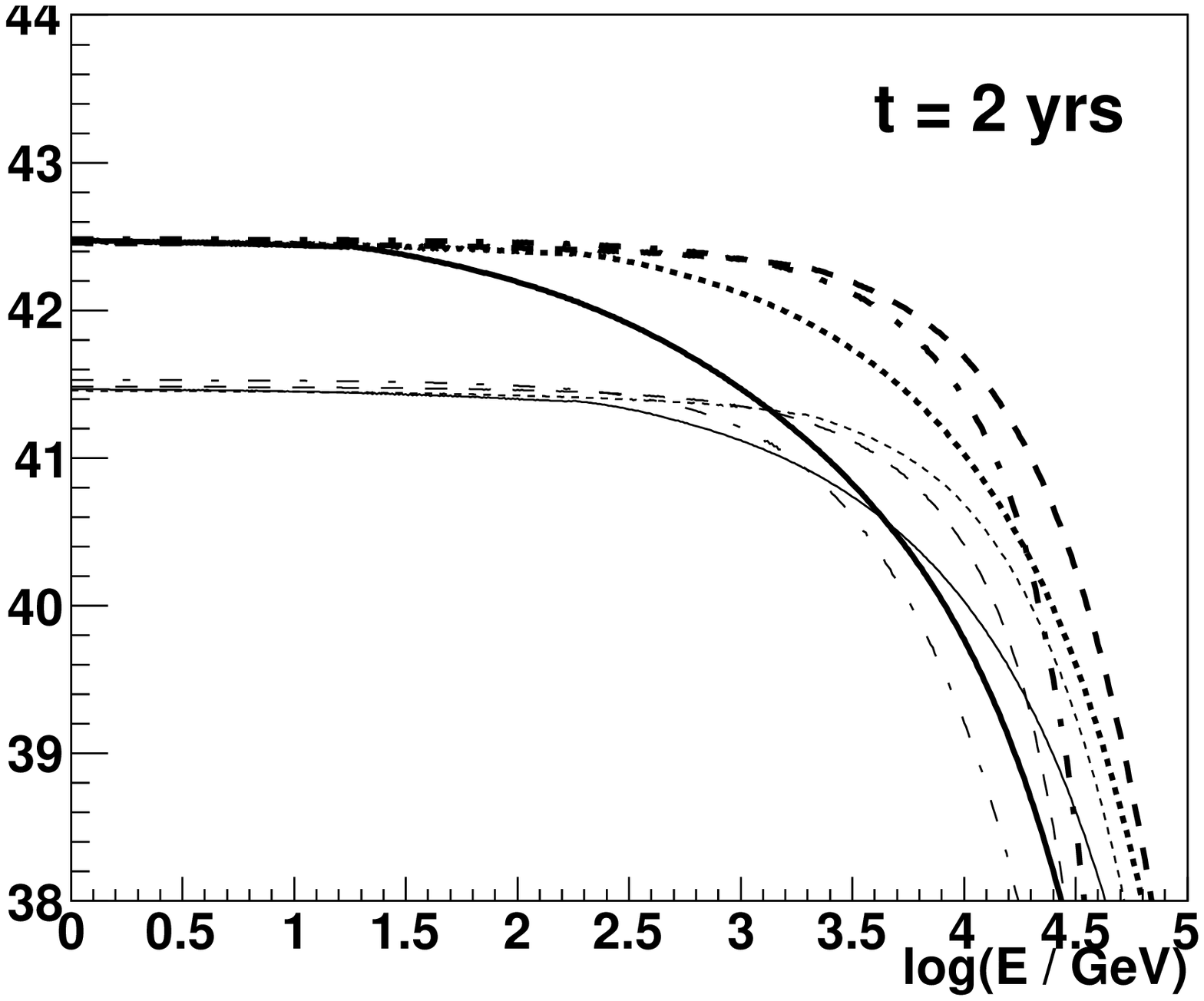}
\includegraphics{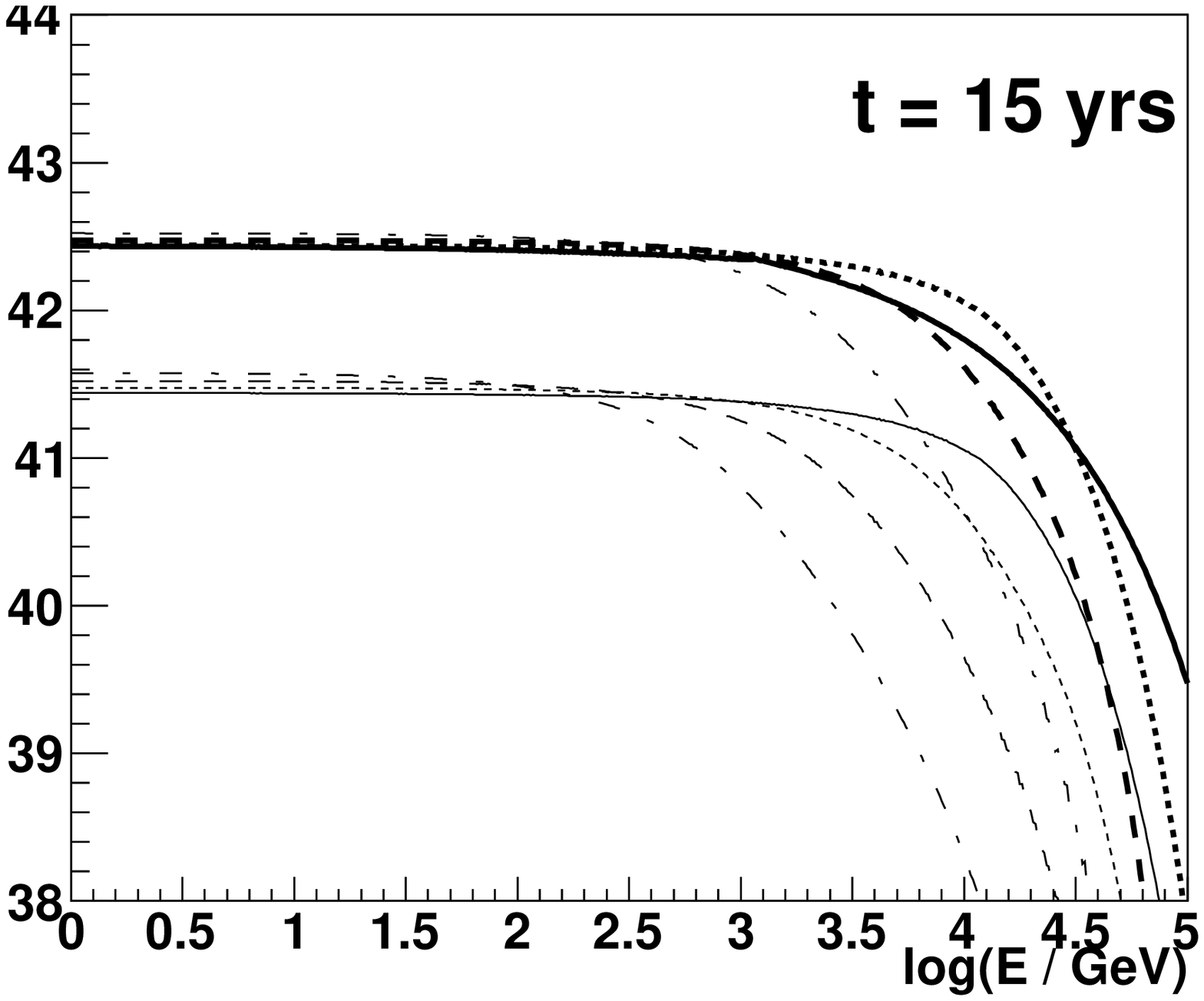}
\caption{The equilibrium spectra of electrons which are accelerated in the shell of Nova for different time 
after the nova explosion:
$t_{\rm obs} = 3$ days (left figure), 1 yr (left-centre), 2 yrs (right-centre), and 15 yrs (right).
Electrons are accelerated continuously during the period from the explosion up to $t_{\rm obs}$ 
accumulating $10\%$ of the kinetic energy of the shell, $E_{\rm sh} = 0.5M_{\rm sh}v_{\rm sh}^2$, 
where the mass of the shell is $M_{\rm sh} = 10^{-6}$~M$_\odot$ (thick curves) and $10^{-7}$~M$_\odot$ (thin).
The velocity of the shell is $v_{\rm sh} = 6\times 10^8v_8$~cm~s$^{-1}$. In all calculations the thickness of the 
shell is $\beta = 0.1$.
Electrons are injected with the power law spectrum with the spectral index 2, with an exponential cut-off 
at $E_{\rm max}$, where $E_{\rm max}$ is calculated from the comparison of the acceleration time scale 
with different energy loss time scales of electrons or with the dynamical time scale of the shell. 
The magnetic field strength at the shell is calculated assuming the equipartition of the magnetic field energy 
density with the kinetic energy density of the shell. The results, for different values of the equipartition 
coefficient $\alpha = 0.1$ (solid curves), $10^{-2}$ (dotted), $10^{-3}$ (dashed), and $10^{-4}$ (dot-dashed) 
are reported.  The energy loss processes of electrons are taken into account as considered
in Sect.~4.}
\label{fig4}
\end{figure*}

Electrons are injected locally (at a specific distance R) with the power law spectrum and the
spectral index equal to 2, with an exponential cut-off at characteristic energy $E_{\rm max}$. 
The spectrum of electrons, $dN/dE/dt = A E^{-2} \exp(-E/E_{\rm max})$, is normalized to their energy gain 
rate from the acceleration
mechanism in the way that the total energy in relativistic electrons, injected from the beginning of the 
explosion up to 
the moment described by the time $t_{\rm max}$, is equal to $10\%$ of the kinetic energy of the shell.
The kinetic energy of the shell is determined by $E_{\rm sh} = 0.5M_{\rm sh}v_{\rm sh}^2$.

Then, the injection rate of electrons, as a function of distance from the central engine, is obtained from
\begin{eqnarray}
{{dN}\over{dEdR}} = {{1}\over{v_{\rm sh}}}{{dN}\over{dEdt}}.
\label{eq12}
\end{eqnarray}  
Electrons, injected at a specific distance, suffer energy losses on different radiation processes which 
importance depends on physical conditions in the shell at specific distance from the Nova progenitor. 
Those physical conditions are defined above.
We use a step space (and also time) method, in steps of $\Delta R$ (and corresponding step time 
$\Delta t = \Delta R/V_{\rm sh}$), in order to determine the modification
of the electron's spectrum during propagation of the shell. The energies of electrons, at the n-th step and the 
propagation time $t_{\rm n}$ of the shell, are obtained from $E_{n} = E_{(n-1)} - \Delta E$, 
where $\Delta E = \Delta t * \dot{E}_{\rm tot}$, where $\dot{E}_{\rm tot}$ is the sum of the energy losses 
of electrons at the time $t_n$ on important energy loss processes, e.g. the IC on the RG radiation field and 
on the photosphere radiation field, on the bremsstrahlung process, and on the synchrotron process. 
The number of relativistic electrons, inside the shell, slowly builds up during the propagation of the Nova shell. 
However, energies of electrons drop due to their energy losses. At the every time step, also freshly accelerated 
electrons are added to the equilibrium spectrum of electrons calculated in the previous step.

Based on the above procedure, we calculate the equilibrium spectrum of electrons at an arbitrary moment
after Nova explosion, corresponding to the specific distance of the shell from the WD.
The example equilibrium spectra of electrons, at a specific time after explosion, equal to 3 days, 1 yr, 2yrs 
and 15 yrs, are shown in Fig.~4. The spectra are shown for two values of the initial mass of the shell, 
$M_{\rm sh} = 10^{-6}$~M$_\odot$ (thick curves) and  $10^{-7}$~M$_\odot$ (thin), and the shell velocity 
$v_{\rm sh} = 6\times 10^8$~cm~s$^{-1}$. 
We investigate the dependence of the equilibrium spectrum of electrons on the magnetic field strength, which 
determines the acceleration process of electrons. The magnetic field strength is also responsible for the main 
energy loss mechanism. It is described by the equipartition coefficient $\alpha$. 
We show that, within a few days after explosion, the equilibrium spectra of electrons are steep since electrons
are able to cool efficiently. This is due to 
the fact that soon after explosion, the radiative energy loss time scales of electrons are shorter than 
the dynamical time scale of the shell. Note that, these spectra can still extend up 
to $\sim$TeV energies. 
Note that, at latter stages of the shell propagation, the cut-off in the electron's spectrum does not depend
strongly on the value of the parameter $\alpha$. This is due to a relatively weak dependence of the maximum energies
of injected electrons on $\alpha$ (see Eqs.~6 and~7, and also Fig.~3).
We observe interesting behaviour of the electron spectra on the mass of the shell. 
Since the magnetic field is assumed to be at some level of the equipartition with the kinetic energy of the shell,
the larger mass means also stronger magnetic field and more efficient cooling of injected electrons.
Therefore, the equilibrium spectra of electrons are steeper, showing the cut-off at lower energies for larger masses 
of the Nova shell.
However, a year after explosion the synchrotron cooling of electrons becomes limited by the dynamical time scale
of the shell. Then, the spectra of electrons are well described by  the power law function with the maximum 
energy extending up to a few tens of TeV. Therefore, we expect that the acceleration process of electrons 
in the Nova shell at the late time after explosion can be tested at TeV energies by the Cherenkov telescopes.

\section{Gamma-rays from the freely expending shell}

In the previous section we have calculated the equilibrium spectrum of electrons at an arbitrary time
after the Nova explosion assuming that those electrons are confined within the freely expending shell of the Nova.
As we have noted above, the medium, in which the Nova shell propagates, is expected to be strongly inhomogeneous.
In general, we distinguished two regions with different proprieties, in the equatorial region of the binary 
system and in its polar region. We apply the freely expending shell model to describe
propagation of the shell in the polar region. 

\begin{figure*}
\vskip 4.5truecm
\includegraphics{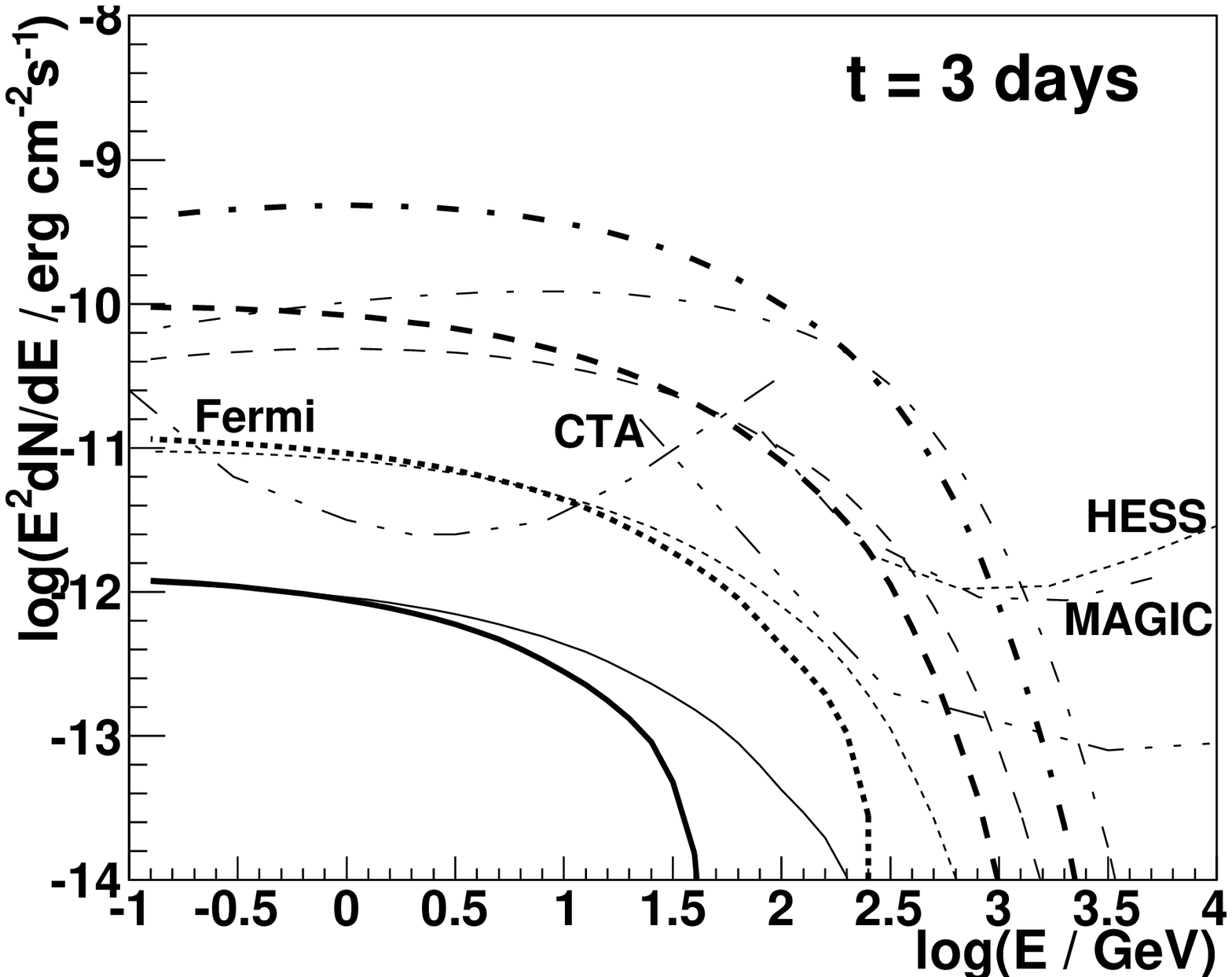}
\includegraphics{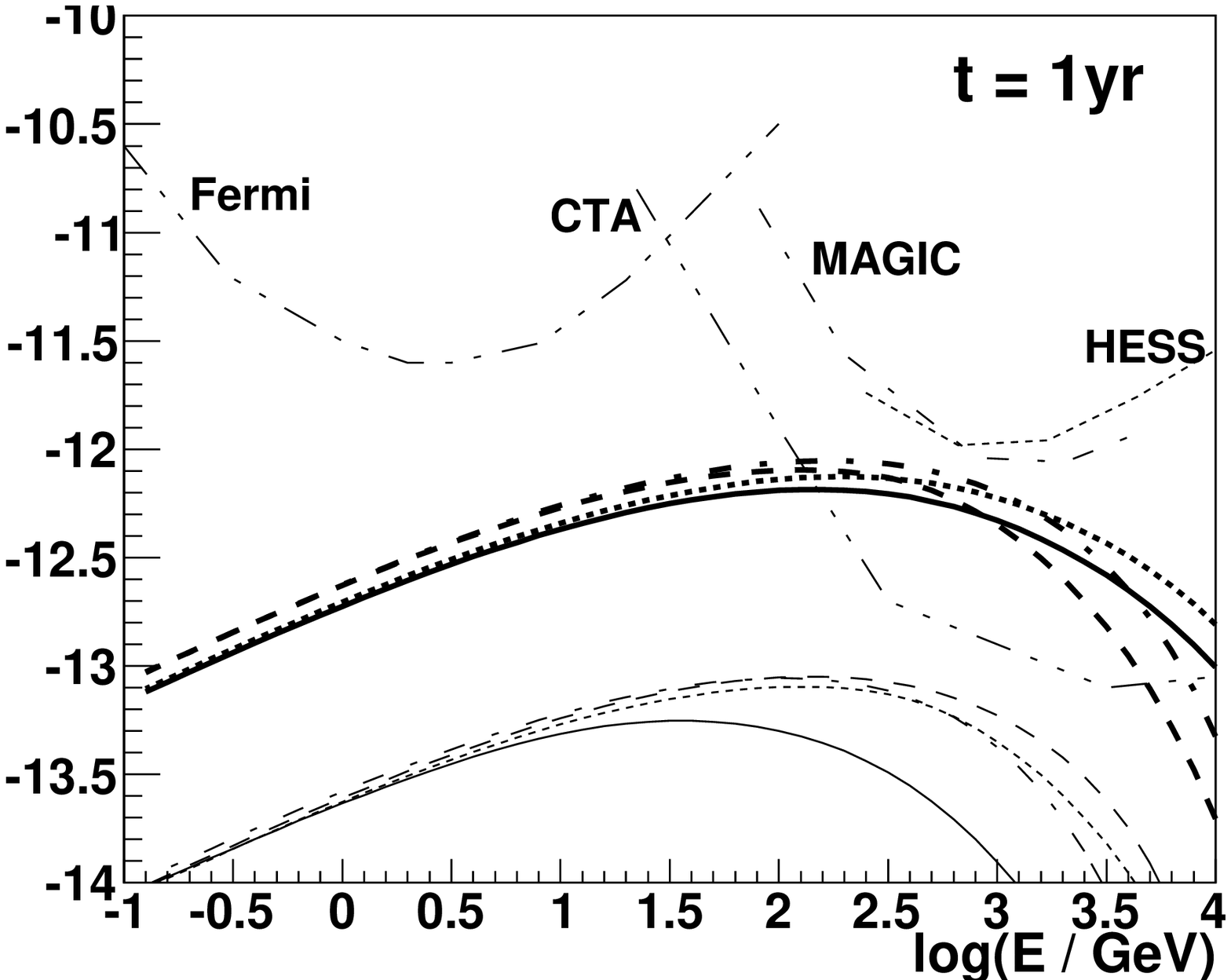}
\includegraphics{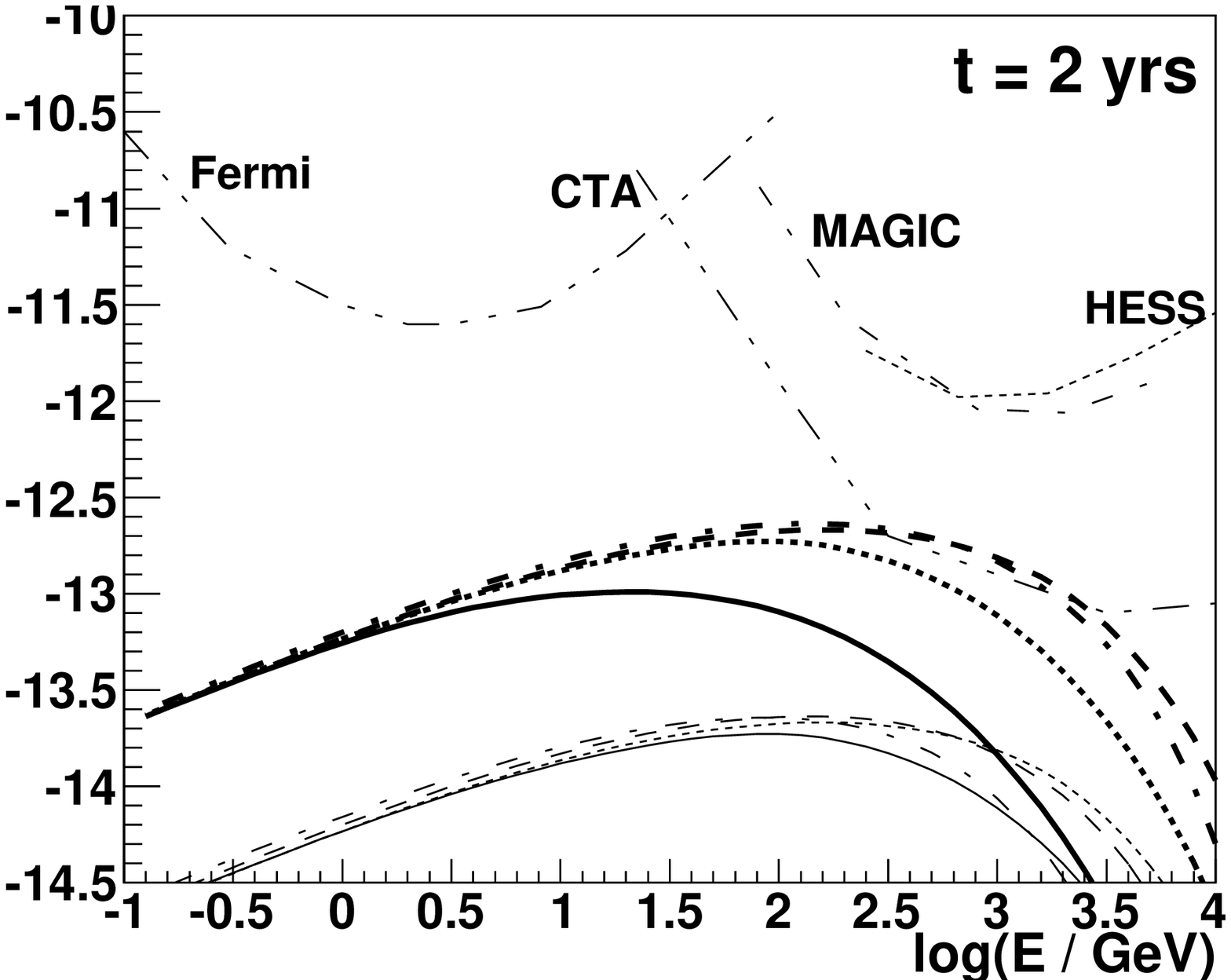}
\includegraphics{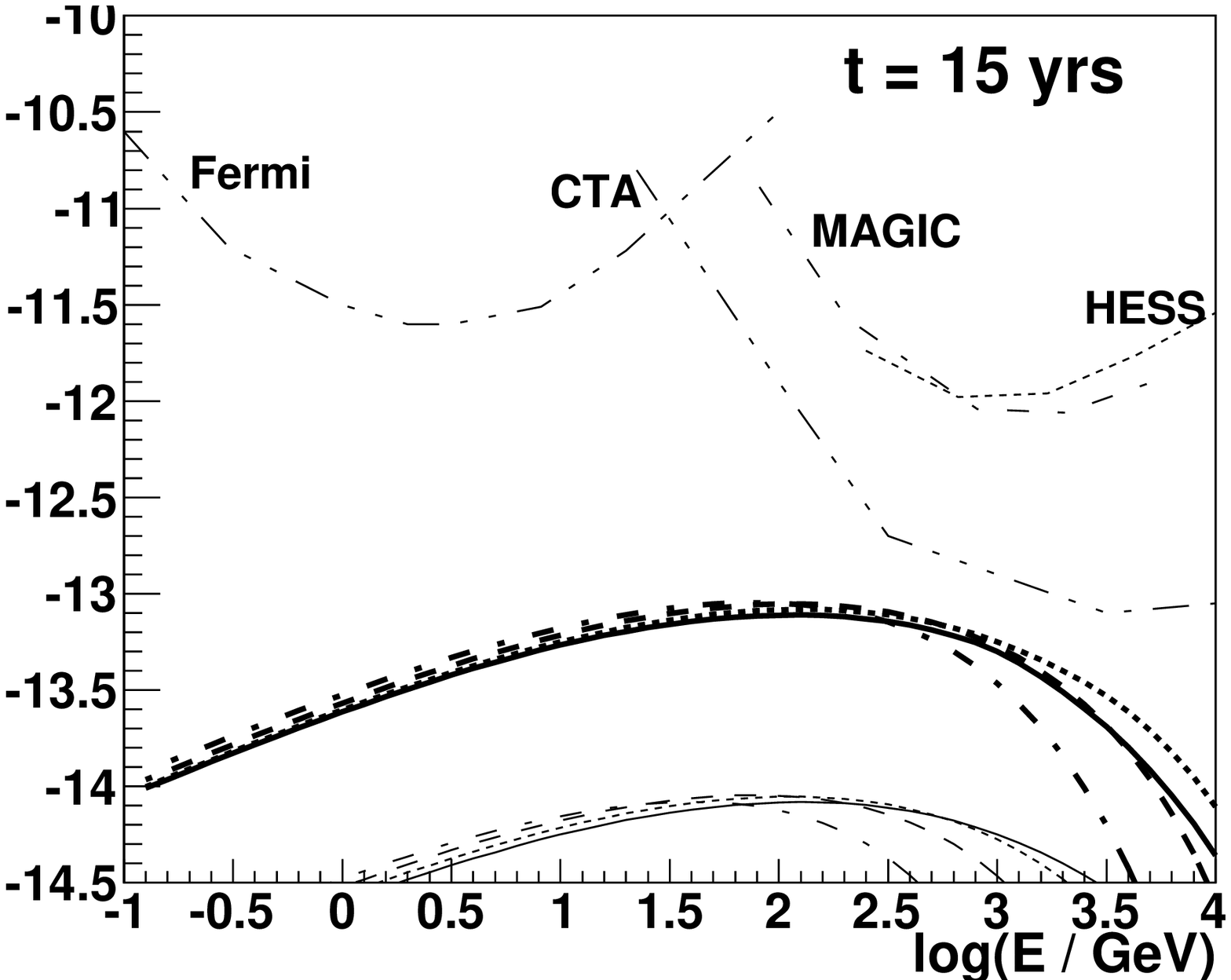}
\caption{The $\gamma$-ray spectra from the IC scattering of the soft radiation from the Nova photosphere 
and the Red Giant by the equilibrium spectrum of electrons calculated at different moments after the initial 
Nova explosion: $t_{\rm obs} = 3$ days (figure on the left), 1 yr (centre-left), 2 yrs (centre-right),
and 15 yrs (right). In the case of 3 days and 15 yrs, the soft radiation from the Nova photosphere dominates. 
Electrons are continuously ejected with the power law spectrum with the spectral index 2 up to 
$E_{\rm max}$, collecting $10\%$ of the kinetic energy of the shell. The magnetic field in the shell
is estimated from the equipartition of the magnetic field energy density with the kinetic energy density of the shell 
with the equipartition coefficient assumed to be 
$\alpha = 0.1$ (solid curve), $10^{-2}$ (dotted), $10^{-3}$ (dashed), and $10^{-4}$ (dot-dashed). 
The other parameters of the model are these same as in Fig.~4. For comparison we show the sensitivities of the
Fermi-LAT (100 days, Funk et al.~2013),
the MAGIC  Cherenkov telescopes (50 hr sensitivity, marked by the dot-dashed curve, Aleksi\^c~et al.~2012), the HESS (25 hrs sensitivity, marked by the dotted curve, see HESS Collaboration),
and the Cherenkov Telescope Array (50hr, Maier et al.~2017).
}
\label{fig5}
\end{figure*}

We calculate $\gamma$-ray fluxes from electrons which comptonize the soft radiation produced 
by the Red Giant star in the binary system with the parameters mentioned in the Introduction. 
The distance to the Nova RS Oph is not well known. We assume the
value 2.45 kpc (Rupen et al.~2008, and see discussion in Acciari et al.~2022).  
At the early stage of the Nova explosion, this soft radiation field is in fact dominated by the soft radiation 
from the Nova photosphere (defined in Sect.~4). The photosphere emission forms also dominant radiation
field for
electrons in the shell at the moment of the subsequent explosion of the recurrent Nova. Therefore,
at the early stage of shell propagation, and also at the moment of the next explosion, additional radiation 
from the photosphere of the Nova is taken into account when calculating the $\gamma$-ray spectrum.  
At the early time, electrons in the shell lose efficiently energy on the radiation processes since their 
radiative cooling times are shorter than the dynamical time scale of the shell. However, at latter 
phase, electrons lose only a part of their ejected energy. 
Those electrons are advected with the shell to larger distances. 
They finally escape into the interstellar space when the Nova shell becomes decelerated as a result of the 
entrainment of the interstellar matter.

As an example, we show the $\gamma$-ray spectra from the IC process of electrons at 3 days after initial 
explosion, i.e. corresponding to the moment of the observation of the sub-TeV $\gamma$-ray emission
from the recurrent Nova RS Oph by the MAGIC Cherenkov Observatory (Acciari et al.~2022). 
The features of these spectra are investigated as a function of 
the magnetization of the shell described by the equipartition coefficient $\alpha$.
We show that observed $\gamma$-ray emission from GeV to sub-TeV energies
(Fermi, HESS and MAGIC telescopes, Acciari et al.~2022, Aharonian et al.~2022) is well described for 
reasonable parameters of the model
(see dot-dashed and dashed curves in Fig.~5 for $\alpha\sim 10^{-4} - 10^{-3}$). 
We also calculate the $\gamma$-ray spectra at 1 yr and 2 yrs
after explosion in order to confront them with the future observations with the Cherenkov 
telescopes. We predict that these $\gamma$-ray spectra extend up to the TeV energy
range. They stay below the sensitivity limit of the present Cherenkov telescopes unless the kinetic
energy of the Nova shell is clearly larger than considered in our calculations (the mass of the shell
$M_{\rm sh} = 10^{-6}$~M$_\odot$ and $v_{\rm sh} = 6\times 10^8$~cm~s$^{-1}$). However, we note that 
predicted TeV $\gamma$-ray emission at 1 yr after explosion has a chance to be detected by the future 
Cherenkov Telescope Array (CTA) Observatory. We also calculate the $\gamma$-ray spectra 
15 yrs after initial explosion, i.e. at the moment of the next explosion of the recurrent Nova (RS Oph).
Then, the dominant soft radiation field is provided by the Nova photosphere. In this case,
we predict that the fluxes of $\gamma$-rays, produced in terms of this model, are below sensitivities of 
the present and future Cherenkov telescopes (see Fig.~5).

\section{$\gamma$-rays from decelerated shell}

\begin{figure}
\vskip 10.3truecm
\includegraphics{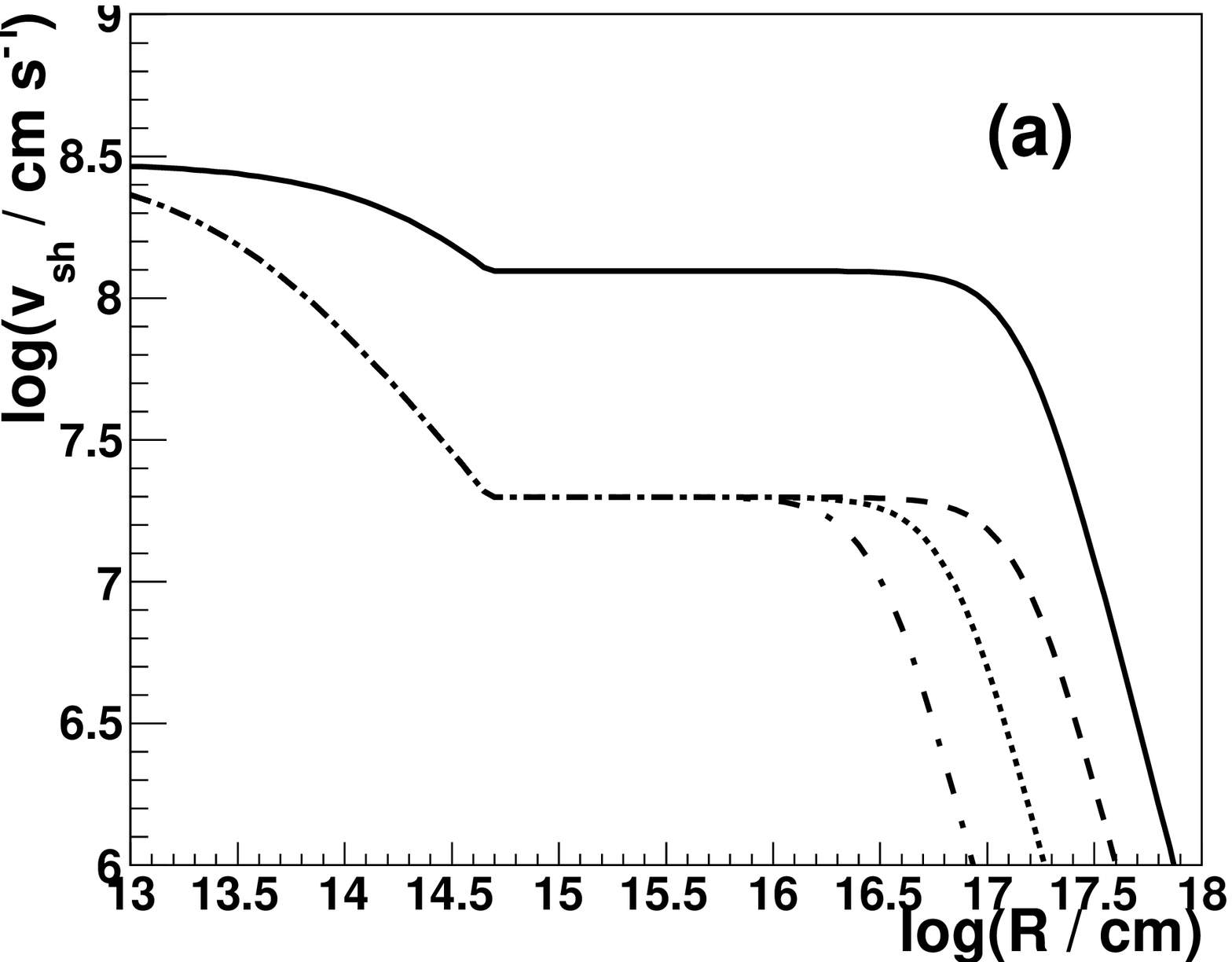}
\includegraphics{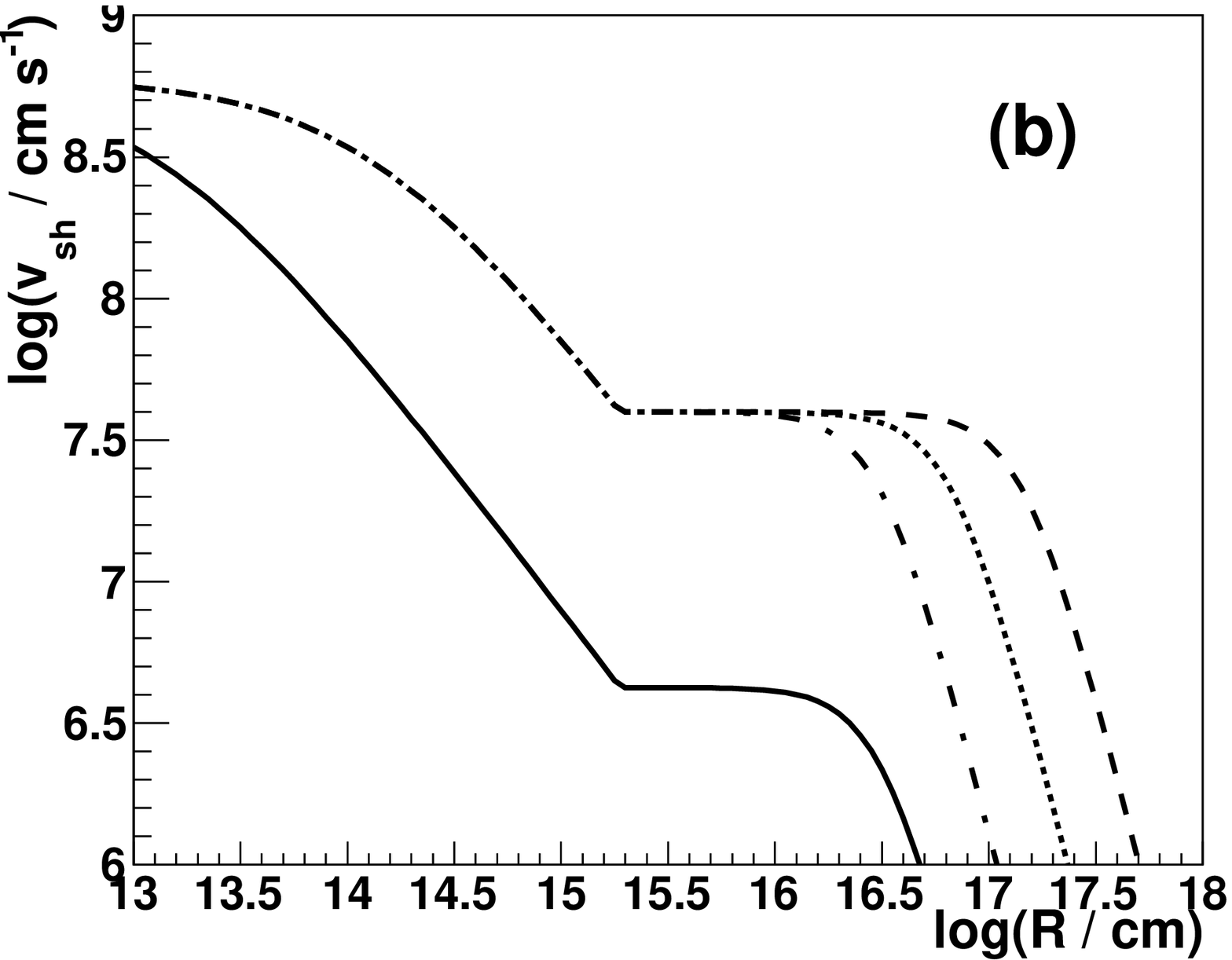}
\caption{The velocity profiles for the shell of the Nova which decelerates due to entrainment of 
the matter from the RG wind and from the surrounding cosmic space. 
(a) The parameters of the shell, the RG 
and the surrounding medium are the following: $M_{\rm sh} = 10^{-5}$~M$_\odot$, 
$v_{\rm sh}^{\rm init} = 3\times 10^8$~cm~s$^{-1}$, $M_{\rm RG} = 10^{-7}$~M$_\odot$~yr$^{-1}$, 
$v_{\rm RG} = 10^6$~cm~s$^{-1}$, $\Omega_{\rm RG} = 0.1$ and $n_{\rm cos} = 1$~part.~cm$^{-3}$ 
(solid curve). 
The profile for $M_{\rm sh} = 10^{-6}$~M$_\odot$ (the RG parameters as above) but for different 
densities of the cosmic medium $n_{\rm cos} = 0.1$~part.~cm$^{-3}$ (dashed), 1~part.~cm$^{-3}$ (dotted) 
and 10~part.~cm$^{-3}$ (dot-dashed). (b) The velocity profiles of the shell for the RG wind with the parameters 
$M_{\rm RG} = 10^{-7}$~M$_\odot$~yr$^{-1}$,  $v_{\rm RG} = 4\times 10^6$~cm~s$^{-1}$, 
$\Omega_{\rm RG} = 0.1$, the mass of the shell $M_{\rm sh} = 10^{-6}$~M$_\odot$ and
its velocity $v_{\rm sh} = 6\times 10^8$~cm~s$^{-1}$,
and $n_{\rm cos} = 10$~part.~cm$^{-3}$ (dot-dashed), 1~part.~cm$^{-3}$ (dotted), and 0.1~part.~cm$^{-3}$ (dashed), 
and $M_{\rm sh} = 10^{-7}$~M$_\odot$ (solid).}
\label{fig6}
\end{figure}

A part of the mass lost by the Red Giant wind is accreted onto the WD and a part is expelled in the form 
of the RG wind. The wind is expected to be limited to the region of the equatorial plane of 
the binary system, which is described by a part of the whole solid angle $\Omega$. Since we assumed 
that the Nova explosion occurs spherically-symmetric, the Nova shell
moves within two regions with different proprieties, i.e. partially within the RG 
wind (in the equatorial region of the binary system) and also partially in the polar region. Therefore, 
the shell structure obtains complicated geometry during its propagation.
In Sect.~5, we considered the polar parts of the Nova shell in which the velocity of the Nova shell is not 
influenced by the RG wind. However, a part of the Nova shell, propagating in 
the equatorial region, is expected to suffer significant deceleration due to the entrainment of 
the RG wind. We can simply estimate the effect of deceleration of the shell by assuming conservation of 
the momentum between the expending shell material and the material entrained from the RG wind. 
Then, simple conservation of the momentum is expressed by, 
$\Omega M_{\rm sh} v_{\rm sh}^{\rm init} = (\Omega M_{\rm sh} + M_{\rm RG}(R))v_{\rm sh}(R)$, where 
$M_{\rm sh} = 10^{-6}M_{-6}$~M$_\odot$ is the initial mass of the shell,
$v_{\rm sh}^{\rm init}$ is the initial velocity of the shell, and 
$M_{\rm RG}(R) = (R/v_{\rm RG})\dot{M}_{\rm RG}\approx 3\times 10^{-21}M_{-7}R/v_6$~M$_\odot$ is the mass of the RG
wind entrained within the shell.
We assume that the previous shell of the Nova has cleaned the surrounding of the Nova. 
Due to a relatively low velocity, the Red Giant wind can only occupy a small region around the binary system. 
It is estimated on  $R_{\rm RG} = v_{\rm RG} t_{\rm rec}\approx 4.7\times 10^{14}v_6$~cm.  
Then, in the equatorial region of the binary system, closer than $R_{\rm RG}$, 
the velocity of the Nova shell evolves according to 
\begin{eqnarray}
v_{\rm sh}(R) = {{\Omega M_{\rm sh}v_{\rm sh}^{\rm init}}\over{(\Omega M_{\rm sh} + M_{\rm RG}(R))}}\approx 
{{v_{\rm sh}^{\rm init}}\over{[1 + {{3\times 10^{-15}M_{-7}R}\over{\Omega M_{-6}v_6}}]}},
\label{eq13}
\end{eqnarray}
\noindent
where R is in cm.
At distances, $R > R_{\rm RG}$, the RG wind is not present. Then, the Nova shell
moves with the constant velocity equal to $v_{\rm sh}(R_{\rm RG})$. In fact, at very large distances, 
the Nova shell can become again decelerated due to
the entrainment of the matter from the surrounding cosmic space.
For the average density of matter in the cosmic space, of the order of $n_{\rm cos} = 0.1n_{0.1}$ 
particles~cm$^{-3}$, this additional 
deceleration is expected to become important on a sub-parsec distance scale. We include this additional 
effect on the shell velocity model introducing additional deceleration factor,
\begin{eqnarray}
v(R) = {{M_{\rm sh}v_{\rm sh}^{\rm init}}\over{[M_{\rm sh} + M_{\rm cos}(R)]}}\approx 
{{v_{\rm sh}^{\rm init}}\over{[1 + 0.3R_{17}^3n_{0.1}/M_{-6}]}},
\label{eq14}
\end{eqnarray}
\noindent
where $v_{\rm sh}^{\rm init}$ is the initial velocity of the shell but already after the process of 
the interaction with the RG wind,
$M_{\rm cos} = (4/3)\pi R^3n_{\rm cos} m_{\rm p}\approx 3\times 
10^{-6}R_{17}^3n_{0.1}$~M$_\odot$ is the amount of the  mass entrained by the shell from the cosmic space,
and the distance of the shell from the Nova progenitor is $R = 10^{17}R_{17}$ cm.
As a result of the above processes, the velocity profile of the shell, propagating in the equatorial region
of the binary system, has complicated dependence on the distance from the WD. The example profiles of 
the shell velocity are shown in Fig.~6. The typical parameters of 
the RG wind, and the surrounding medium, are considered. 
Depending on the parameters of the Nova shell, its mass, initial velocity, and the solid angle of the 
Red Giant wind, we observe moderate or very effective deceleration of the shell 
(see velocity profiles in Fig.~6). We conclude that, a part of the Nova shell, propagating in the equatorial 
region of the binary system, stays much closer to the RG (the main source of soft radiation)
then the freely expending part of the shell towards the polar region.
It is expected that, due to the above mentioned effect, the $\gamma$-ray fluxes from the IC process of electrons
in the equatorial part of the Nova shell reach clearly larger levels.

As in Sect.~5, we calculate the equilibrium spectra of electrons within the Nova shell applying those 
same assumptions on the acceleration model of electrons during the shell propagation. In these calculations, 
we take into account the dependence of the shell velocity on the distance from the central engine of the Nova.
The example calculations
of those spectra are shown on Fig.~7. The equilibrium spectra of electrons, in the decelerating shell model, 
extends to lower energies than observed in the case of freely expending shell model. This is due to stronger
magnetic fields in the shell region at a specific moment of propagation of the Nova shell. However, the equilibrium 
spectra of electrons, in the decelerating shell model, still extend to a few TeV.

\begin{figure*}
\vskip 5.2truecm
\includegraphics{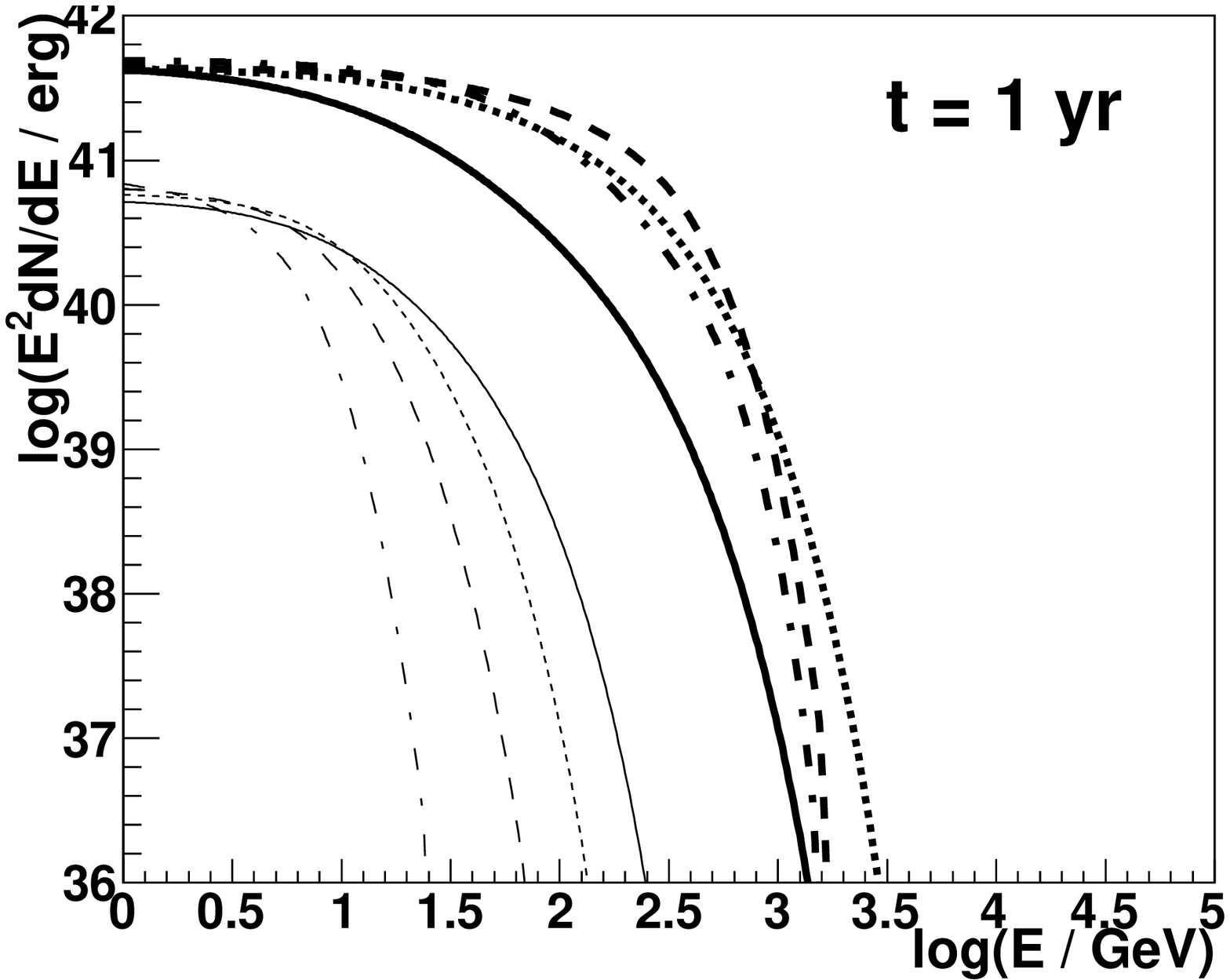}
\includegraphics{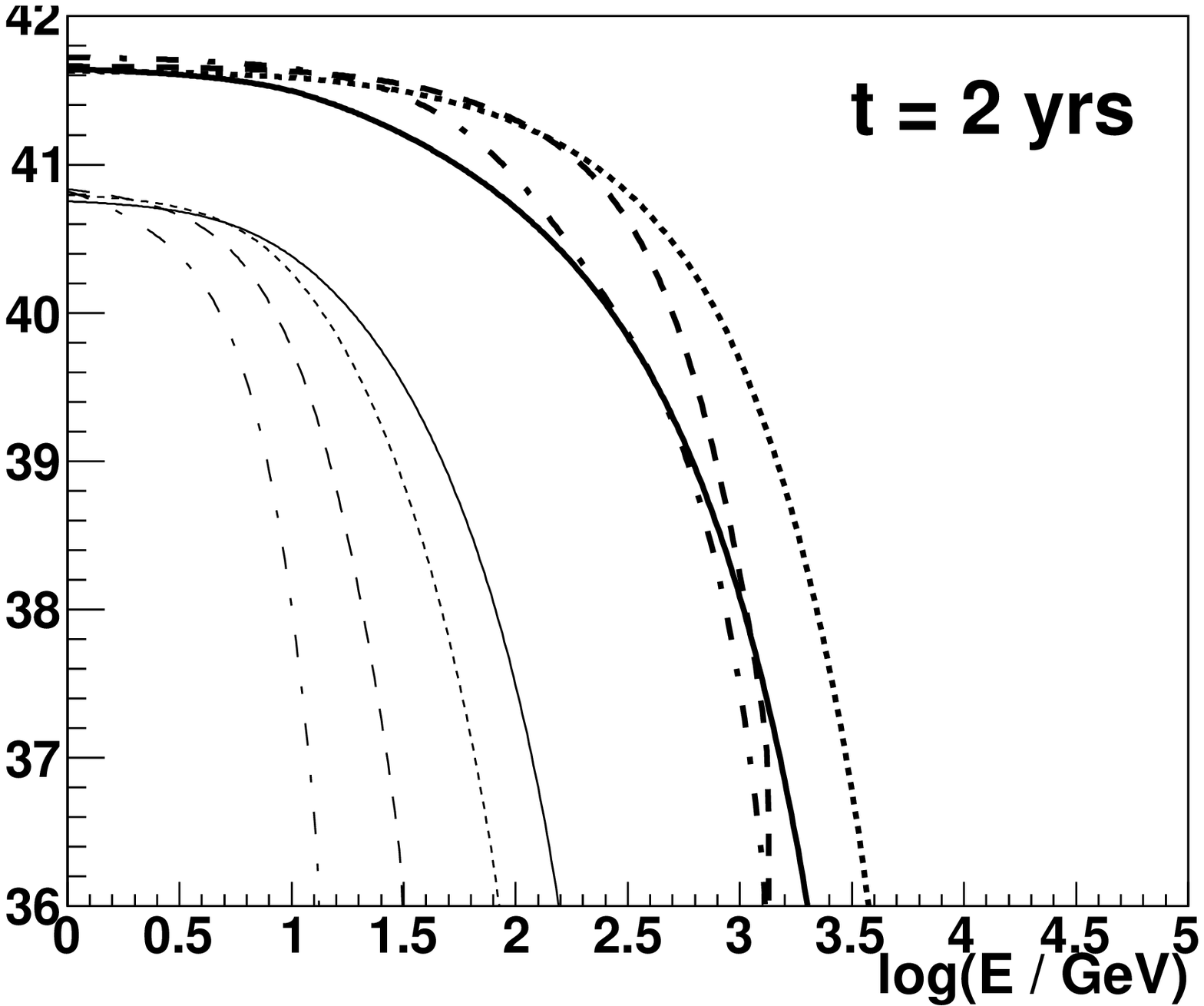}
\includegraphics{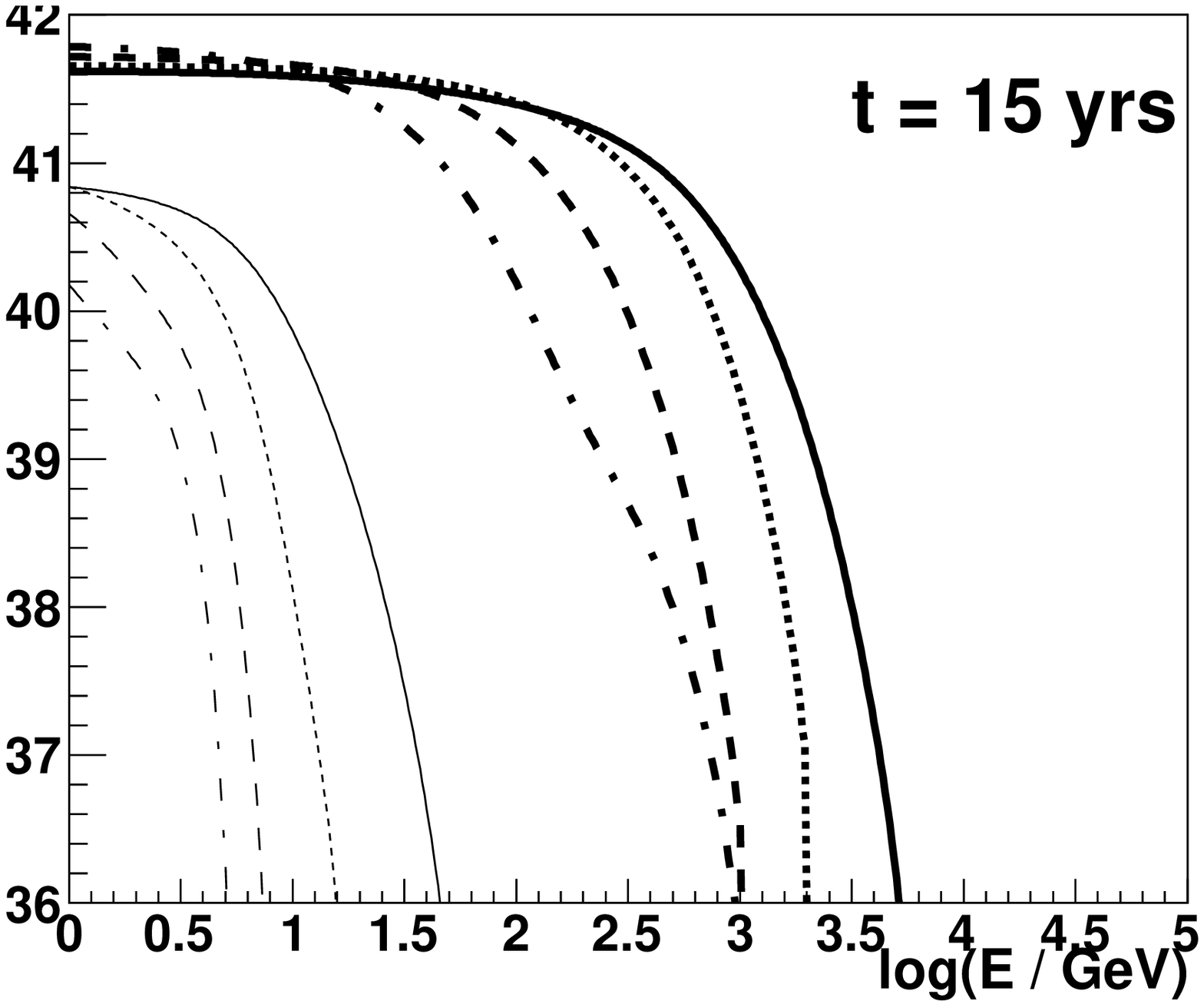}
\caption{As in Fig.~4 but for the the shell, which is decelerated due to the interaction with the wind 
from the RG companion. The model for the shell deceleration is defined in Sect.~7. The electrons, 
with the equilibrium spectrum, up-scatter the soft radiation from the RG at 1 yr (left figure) and 
2 years (centre) after explosion. The radiation from the Nova photosphere, produced during the subsequent flare, 
dominates over 
the radiation from the RG at 15 yrs after the Nova explosion. It is assumed that the RG wind 
is confined within the $\Omega_{\rm RG} = 0.3$ of the solid angle in the equatorial region of the binary 
system. The mass loss rate of the RG is $\dot{M}_{\rm RG} = 
10^{-7}$~M$_{\odot}$~yr$^{-1}$ and the RG wind velocity is 40 km~s$^{-1}$. 
The specific $\gamma$-ray spectra are calculated for the coefficient $\alpha = 0.1$ (solid curve), 
$10^{-2}$ (dotted), $10^{-3}$ (dashed), and $10^{-4}$ (dot-dashed).}
\label{fig7}
\end{figure*}
\begin{figure*}
\vskip 5.2truecm
\includegraphics{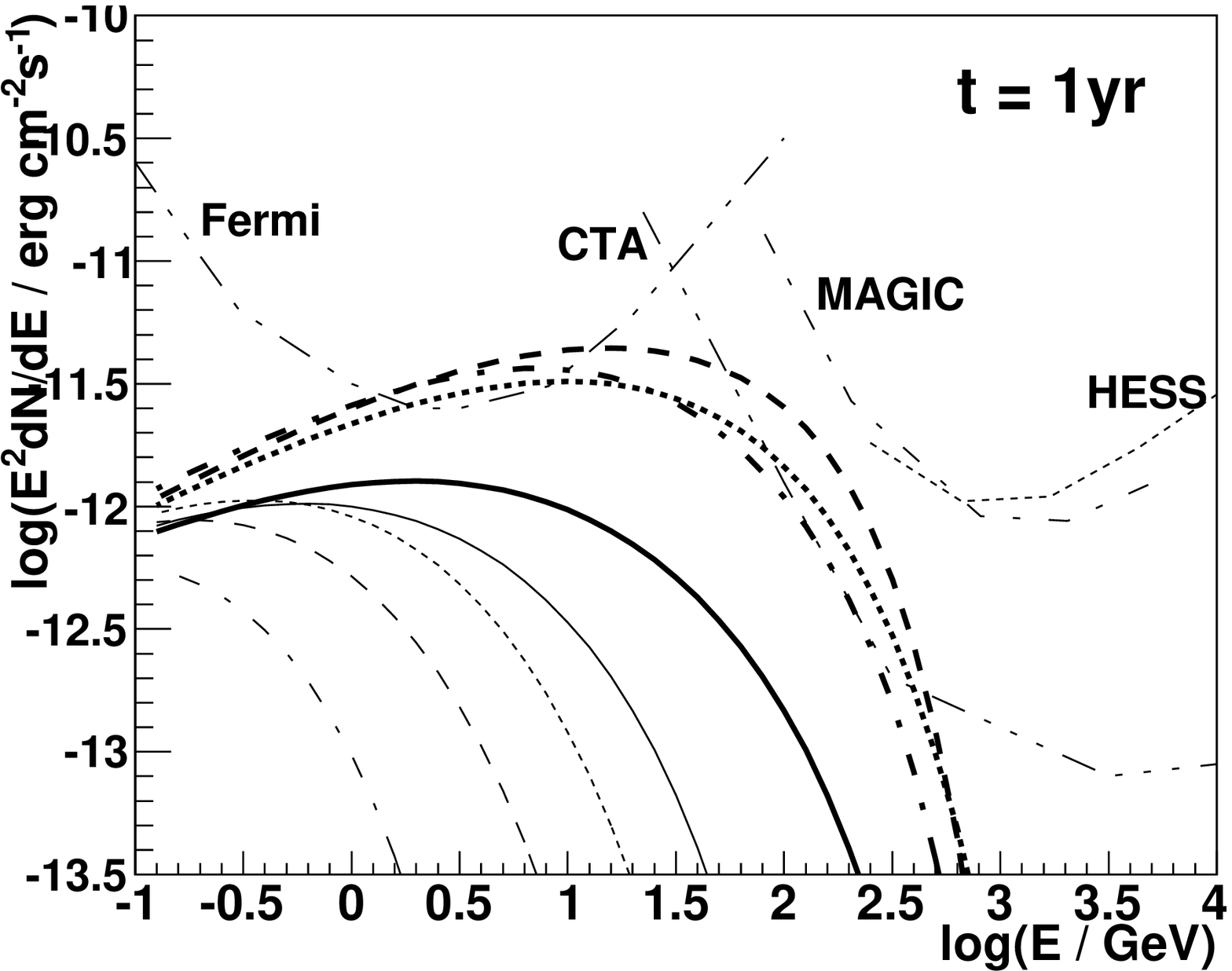}
\includegraphics{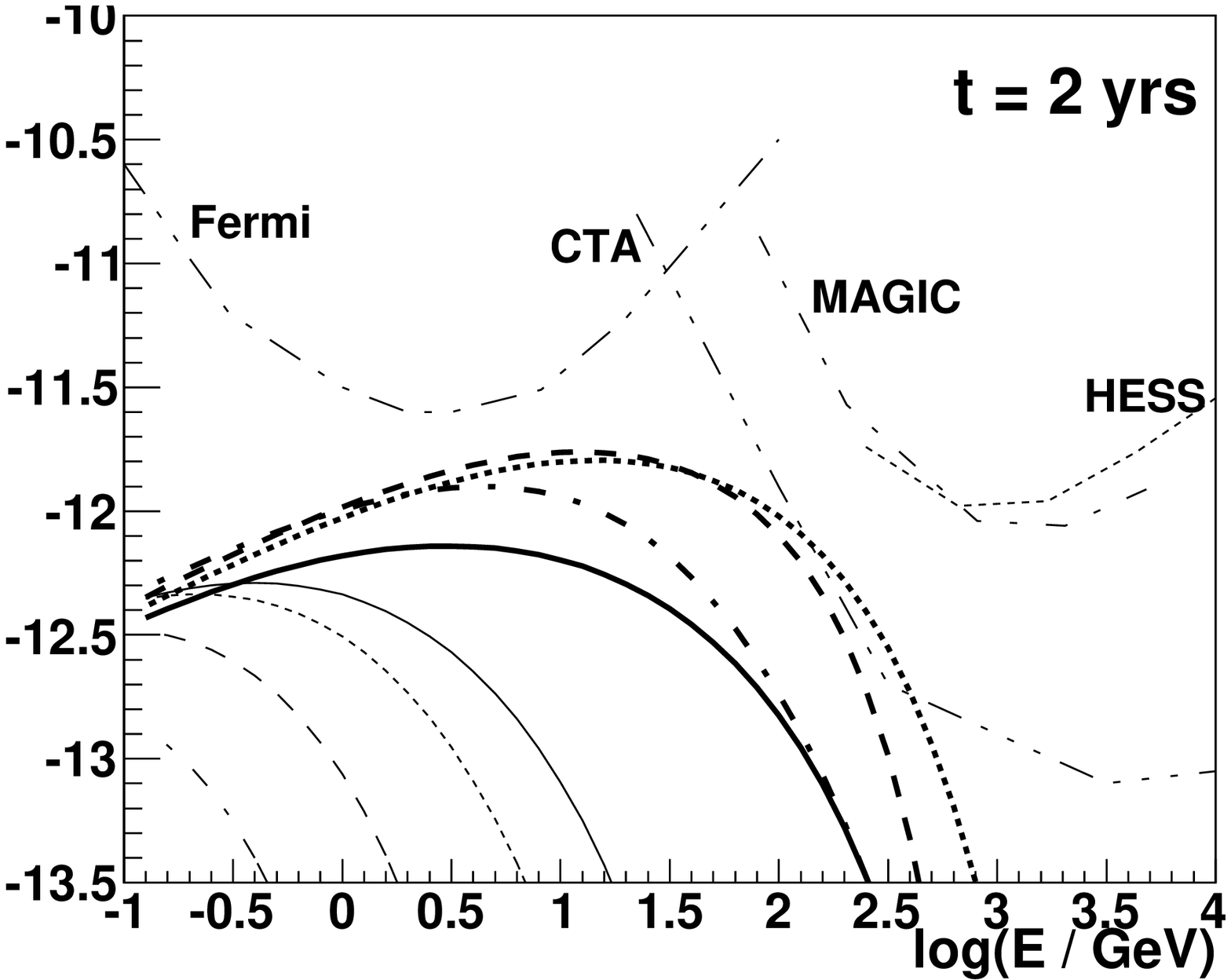}
\includegraphics{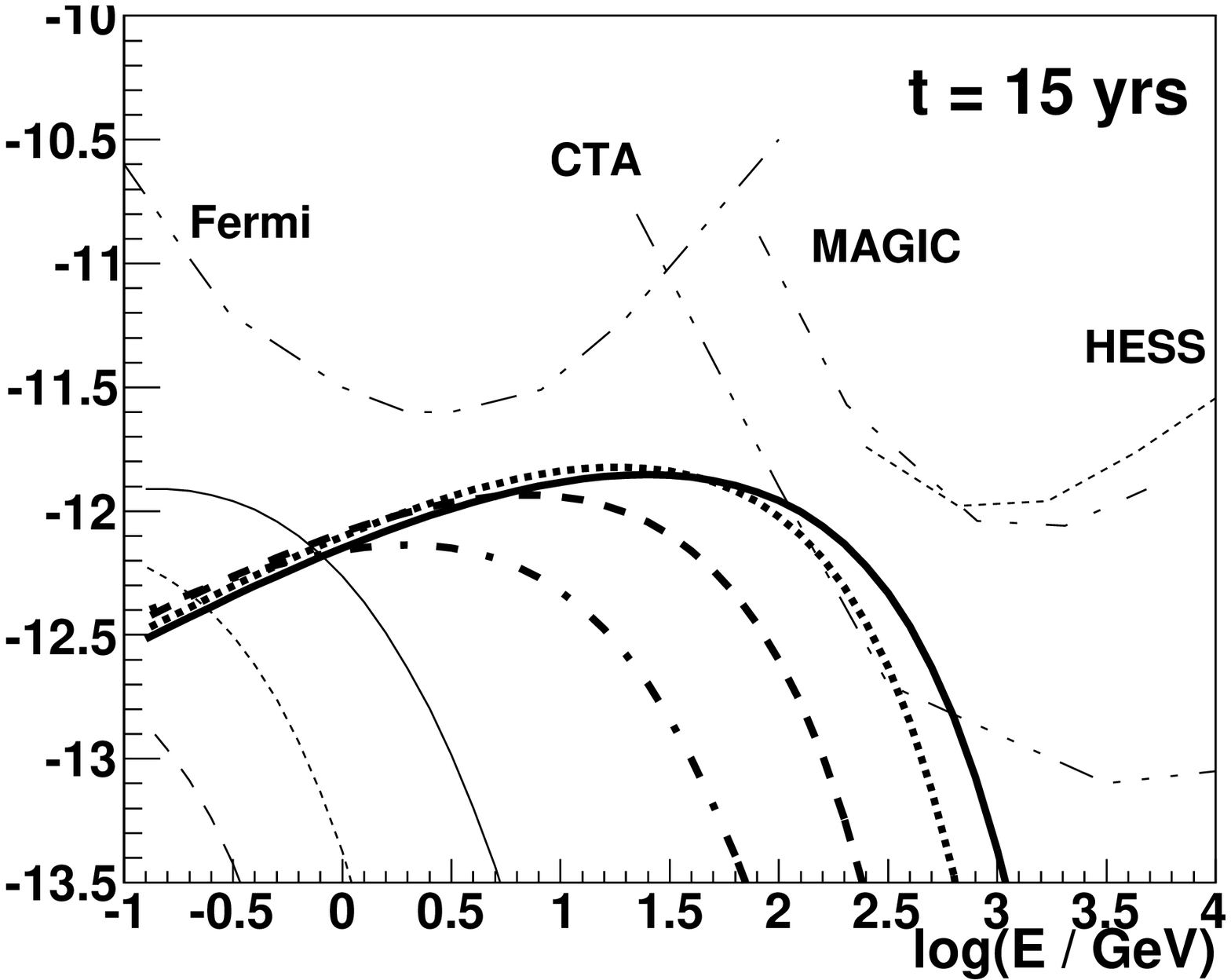}
\caption{As in Fig.~5 but for decelerating Nova shell. The $\gamma$-ray spectra are calculated 
for the parameters of the equilibrium spectrum of electrons as shown in Fig.~7.}
\label{fig8}
\end{figure*}

In Fig.~8, we also calculate the $\gamma$-ray spectra from the IC process for the equilibrium spectra of electrons. In contrast to the model of free expending Nova shell, the $\gamma$-ray fluxes, for
these same parameters as considered in the case of freely expending shell, i.e.
strongly magnetised shells with the mass equal to $M_{\rm sh} = 10^{-6}$~M$_\odot$ and the energy conversion
efficiency from the shell to relativistic electrons equal to $10\%$, 
are expected to be within the sensitivity limit of the CTA. At 1 year after the Nova explosion and large masses of the Nova shell, the 
$\gamma$-ray fluxes are expected to be still close to the 50 hrs sensitivity limits of the present Cherenkov 
telescopes (such as HESS, MAGIC and VERITAS). This $\gamma$-ray emission have a chance  to  be
also constrained by the future observations in the GeV energy range by the Fermi Observatory.
We conclude that observations of the GeV-TeV $\gamma$-ray emission from late stages of Nova explosions
should provide useful constraints on the acceleration of leptons in the Nova shells.

\begin{figure*}
\vskip 5.2truecm
\includegraphics{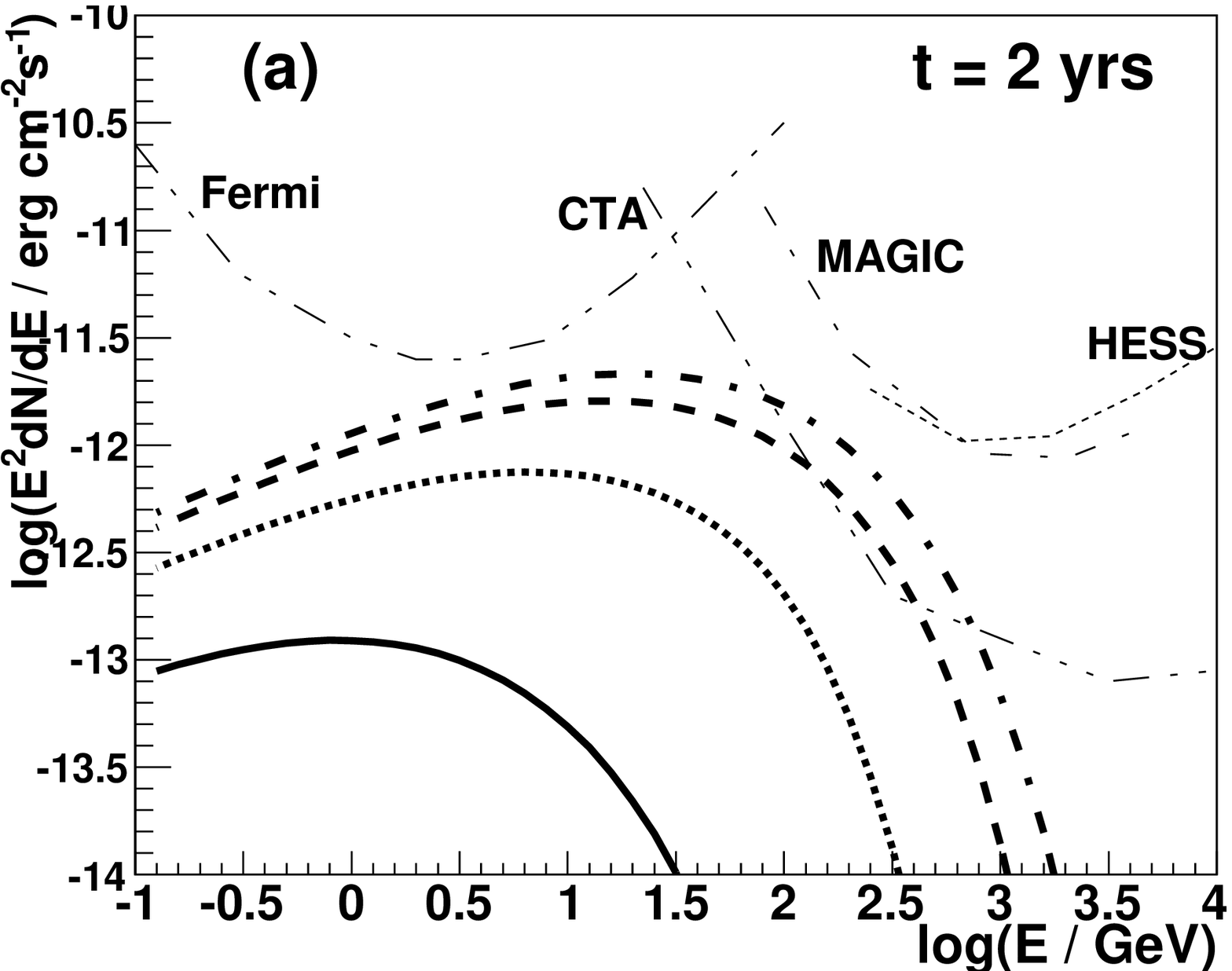}
\includegraphics{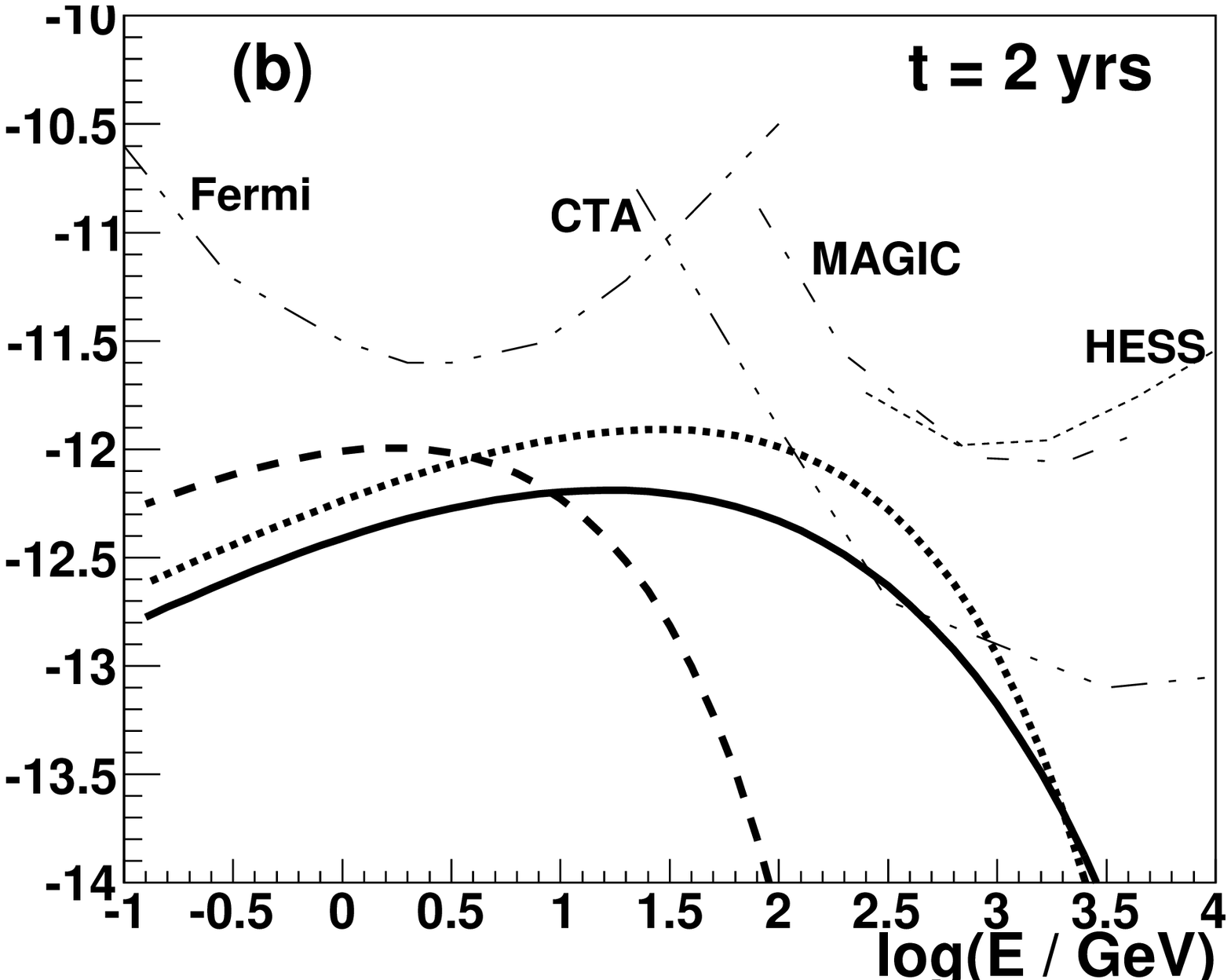}
\includegraphics{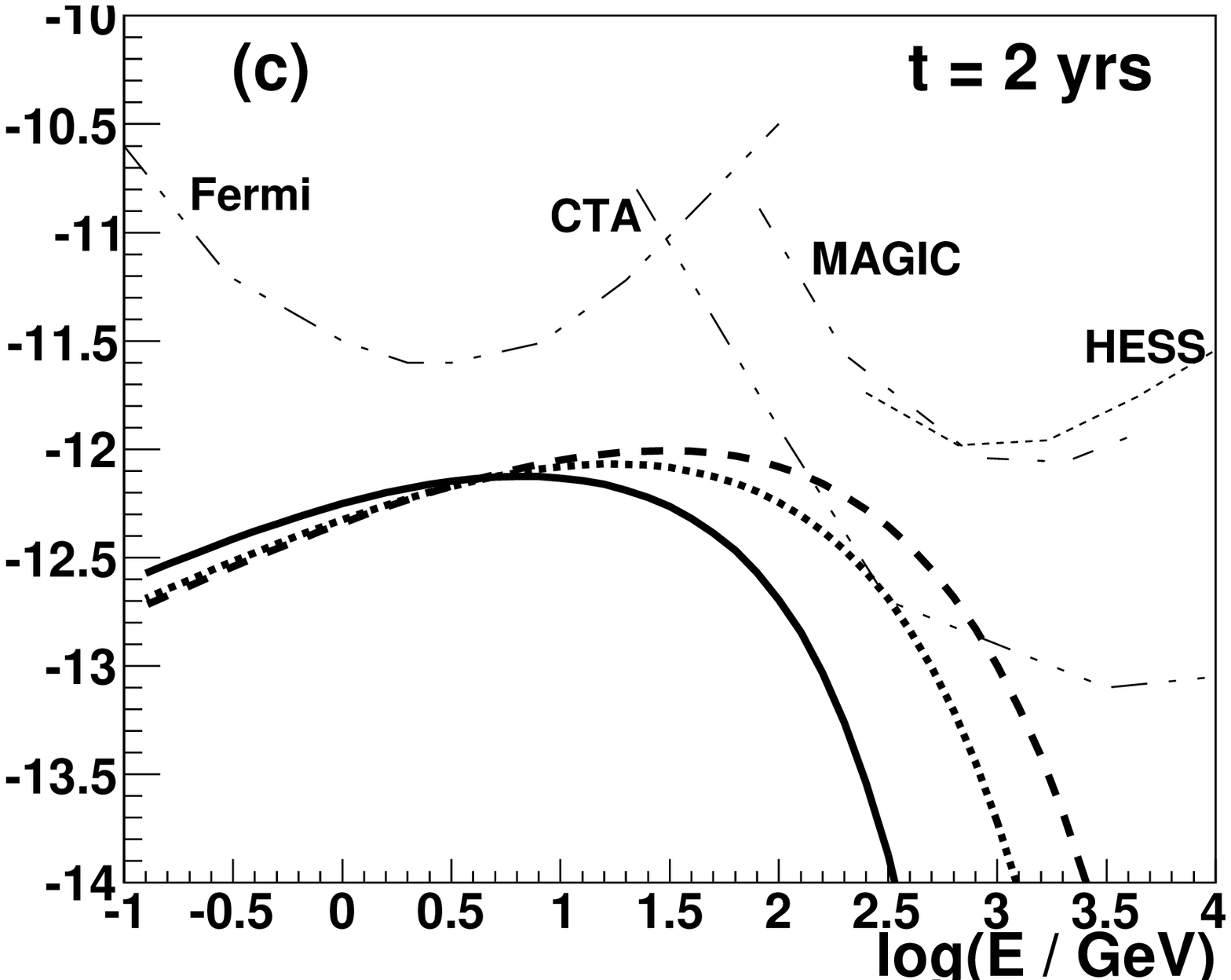}
\includegraphics{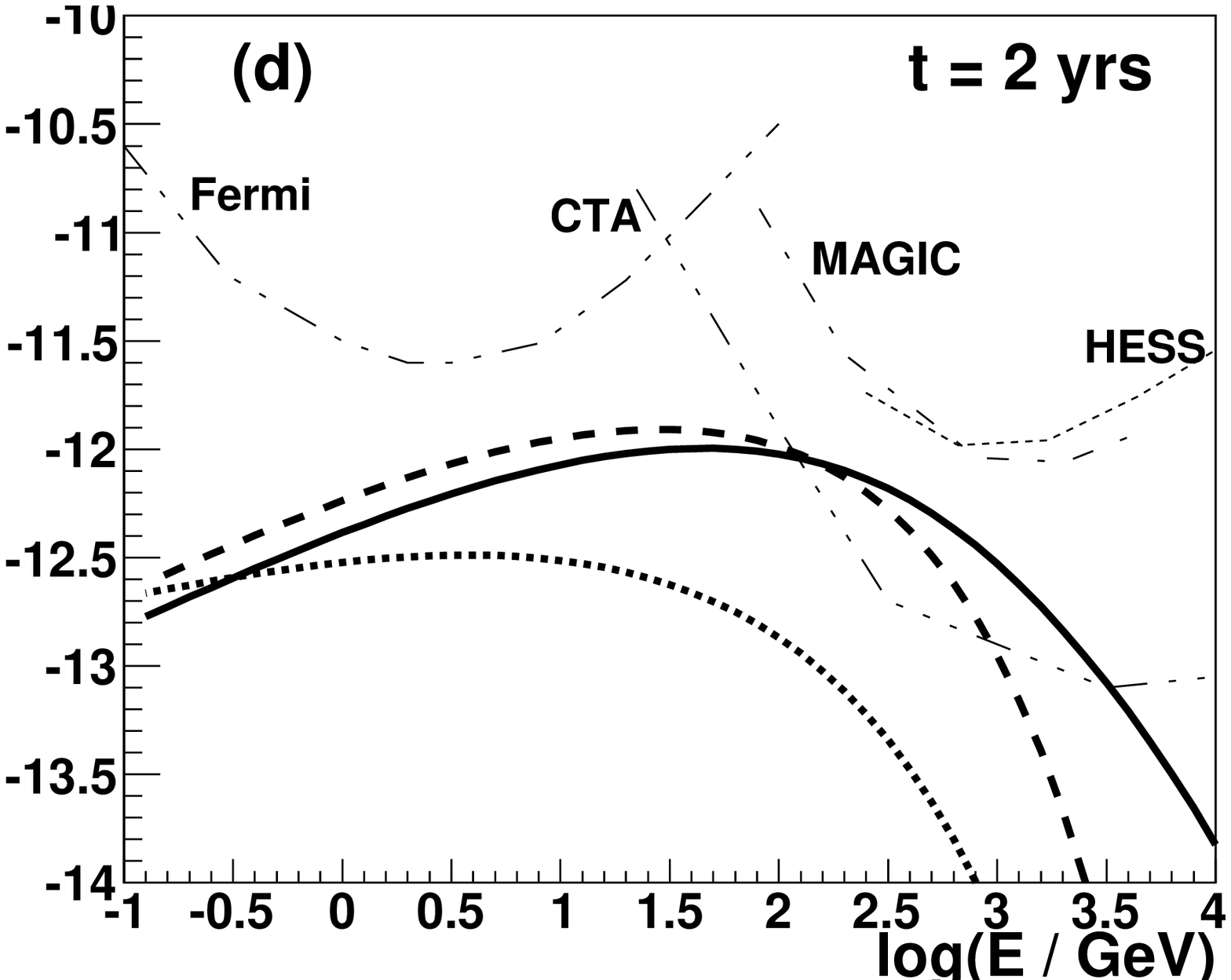}
\caption{The $\gamma$-ray spectra produced in terms of the decelerated shell model, as considered in 
Figs.~7 and 8, at 2 yrs after initial explosion of Nova. 
(a) The dependence on the velocity of the shell $v_{\rm sh} = 10^8$~cm~s$^{-1}$ (solid curve),
$3\times 10^8$~cm~s$^{-1}$ (dotted),  $6\times 10^8$~cm~s$^{-1}$ (dashed), $9\times 10^8$~cm~s$^{-1}$ 
(dot-dashed). The mass of the shell is fixed on $M_{\rm sh} = 10^{-6}$~M$_\odot$ and the parameter 
$\alpha = 0.01$.
(b) The dependence on the mass of the expending shell of Nova $M_{\rm sh} = 10^{-5}$~M$_\odot$ (solid), 
$10^{-6}$~M$_\odot$ (dotted), and  $10^{-7}$~M$_\odot$ (dashed). The other parameters are 
$v_{\rm sh} = 6\times 10^8$~cm~s$^{-1}$, $\Omega_{\rm RG} = 0.3$ and $\alpha = 0.01$.
(c) the dependence on the solid angle of the Red Giant wind for $\Omega_{\rm RG} = 0.1$ (solid),
0.5 (dotted), and 0.9 (dashed), for $M_{\rm sh} = 10^{-6}$~M$_\odot$, $v_{\rm sh} = 3\times 10^{8}$~cm~s$^{-1}$ 
and $\alpha = 0.01$. The spectral index of injected electrons is fixed on 2.
(d) Different prescriptions for the acceleration coefficient, $\xi = v_{\rm sh}/c$ (solid) and 
$\xi = 0.1 (v_{\rm sh}/c)^2$ (dashed). The dotted curve shows the $\gamma$-ray spectra for the
spectral index of ejected primary electrons equal to 2.5. The other parameters are 
$v_{\rm sh} = 6\times 10^8$~cm~s$^{-1}$, $\Omega = 0.3$ and $\alpha = 0.01$.}
\label{fig9}
\end{figure*}

We also investigate the dependence of the $\gamma$-ray emission
for the broader range of the model parameters. In Fig.~9a, the $\gamma$-ray spectra are shown for different 
initial velocities of the Nova shell and other parameters fixed to constant values.
The shell with larger velocity produce $\gamma$-ray spectra with larger fluxes. This effect is the result of a 
few counter-working effects. The increase of the fluxes is due to proportionally larger kinetic energy of 
the shell and less efficient acceleration of electrons in a weaker magnetic filed of the shell. 
However, due to larger shell velocity, the shell reaches larger distances from the central engine. 
Then, the soft radiation field, determining the IC process, is reduced.    
In Fig.~9b, the $\gamma$-ray spectra are shown as a function of the mass  of the Nova shell. For the larger mass, 
the Nova shell propagates easily over larger distances since the deceleration process is less significant. 
Then, the magnetic field within the shell should drop to lower values. On the other hand,
larger mass of the shell results in stronger magnetic field due to the equipartition arguments.  
Therefore, the interpretation of the obtained $\gamma$-ray spectra becomes complicated.
The $\gamma$-ray spectra extend to larger energies with the mass of the shell. 
However, if the mass of the shell becomes very large, then it moves to large distances from the source of soft radiation (i.e. the Red Giant). Then, the efficiency of $\gamma$-ray production drops (see dotted and solid curves in Fig.~9b).
In Fig.~9c, we show the dependence 
of the $\gamma$-ray spectra on the solid angle of the RG wind. For larger solid angle, a larger part of 
the shell (and so its kinetic energy) is able to interact with the RG wind. Therefore, the energy available for the
acceleration of electrons becomes larger. From another site, a part of the shell within the equatorial 
region of the RG wind propagates over larger distances during this same time scale, since the amount of the matter
entrained from the Red Giant wind is fixed. This results in 
a weaker magnetic field in the shell region and in consequence larger energies of accelerated 
electrons. Therefore, the $\gamma$-ray spectra extends to larger energies with larger solid angle in which 
the RG wind is confined. 
Finally, in Fig.~9d we investigate the dependence of the $\gamma$-ray spectra on the parameters which determine the acceleration process of electrons. We compare the $\gamma$-ray spectra for two different prescriptions for the acceleration
parameter $\xi$, i.e. depending linearly $\xi = v_{\rm sh}/c$, or quadratically on the shell velocity
(see Eq.~3). It is found that for more efficient acceleration, the $\gamma$-ray spectra do not extend proportionally to the value of $\xi$. This effect is due to the fact that at a few years after explosion the bulk of electrons within the shell reach the maximum energies which are determined by the dynamical time scale of the shell but not by their synchrotron energy losses. 
However, in the case of both prescriptions for the acceleration efficiency, $\gamma$-rays in the TeV energy range
are still expected provided that the parameter $\alpha$ is relatively large ($\alpha = 0.01$). 
From Eq.~7, it is clear that electrons are not able to produce TeV $\gamma$-rays for the case of weakly magnetized shells of Novae 
(i.e. $\alpha << 0.01$).
We also show the $\gamma$-ray spectrum obtained in the case of electrons injected with the steeper power law spectrum then
considered in the above calculations, i.e. spectral index equal to 2.5. As expected, the $\gamma$-ray spectra become 
also steeper in such a case. They might become two weak at TeV energies to be
detected by the Cherenkov telescopes.
We conclude that, at the late stage of the Nova expansion, $\gamma$-ray emission can be easier constrained 
by the $\gamma$-ray telescopes in the case of Novae which shells are moving with larger velocities, having 
larger injection masses, and which equatorial winds of their Red Giants are confined within larger solid angles.
Moreover, the ejection process of electrons should operate for at least a few years and the spectra should be relatively flat, i.e. spectral indexes close to 2, as expected in the case of strong shocks.

\section{Discussion and Conclusion}

We investigate the problem of the time scale for the acceleration of electrons in shells of recurrent Novae.
Recent observations of the $\gamma$-ray emission from several Novae in the GeV energy range, 
and also sub-TeV energies in the case of the Nova RS Oph (see references in the Introduction), 
indicate that such energetic emission
is observed typically up to several (or in some cases to a few tens) of days after initial explosion of Nova.
This is clearly different than observed in the case of supernovae, in which $\gamma$-ray emission 
becomes detectable at latter stages of expansion of the supernova shell. We wonder whether such persistent 
acceleration of particles is also present in the case of Nova shells.
In order to put light on this question, we calculate the $\gamma$-ray IC emission expected from electrons
accelerated in the expending Nova shell at the late time after explosion. 
As an example, we consider in a more detail the case of the Recurrent Nova RS Oph, recently observed 
by the Cherenkov telescopes at a few days after explosion.
A simple, geometrical two component model is elaborated for the expansion of the shell in the medium surrounding 
the Nova. In the equatorial region of the binary system, the shell efficiently decelerates 
due to the interaction with the Red Giant wind. At the polar regions, the shell expends with a constant velocity.  

We investigate a simple time dependent model for the acceleration of electrons and their energy losses.
In terms of such a model, the equilibrium spectra of electrons are calculated at an arbitrary time 
after Nova 
explosion. We show that electrons can be accelerated to TeV energies in the late stage of the Nova.
They produce TeV $\gamma$-rays in the inverse Compton scattering process of the RG soft radiation or
the soft radiation from the Nova photosphere during the time of its domination, i.e at several days after explosion
or at the moment of the next explosion of the recurrent Nova (in the case of RS Oph it is $\sim$15 yrs). 
The IC $\gamma$-ray emission, calculated at a few days after explosion, has the features similar 
to that observed by the Fermi at GeV energies and the Cherenkov telescopes at sub-TeV energies. 

We show that $\gamma$-ray fluxes, produced in terms of such an extended in time acceleration model of electrons,
could be detectable by the planned CTA Observatory for specific conditions in the Nova shell. 
In general, the $\gamma$-ray emission from a part of the shell, propagating within the solid angle of the RG 
wind (equatorial region of the binary system), is more likely to be observed. In fact, this part of the shell 
becomes 
significantly decelerated staying relatively close to the soft radiation from the RG. 
We predict that the CTA should be able to detect TeV $\gamma$-ray emission from the equatorial part of 
the shell of RS Oph even up to the moment of the next explosion (i.e. $\sim$15 yrs), 
provided that the mass of the shell is of the order of $\sim 10^{-6}$~M$_\odot$, the initial velocity of the shell
is $v_{\rm sh} = 6\times 10^8$~cm~s$^{-1}$, and the magnetic field strength in the shell is not very far from the 
equipartition with the kinetic energy of the shell. At the time of $\sim$1 yr after explosion, the predicted 
$\gamma$-ray fluxes are even close to the sensitivity of the present Cherenkov telescopes. Note however, that 
in this case the  
acceleration efficiency of electrons can be also constrained at GeV energies by the future observation 
with the Fermi Observatory.
We also show that $\gamma$-ray fluxes, expected in terms of such acceleration model of electrons 
in shells of Novae, are easier to detect in the case of a more geometrically extended RG wind, 
since in such a case larger part of initially spherical shell of the Nova is decelerated by the RG wind.

Almost freely expanding part of the Nova shell (in the polar regions of the binary system), is expected to 
produce $\gamma$-ray fluxes generally on a lower level than considered above decelerated part of the shell.
This is due to the larger distance of the shell from the soft radiation field of the RG, or from 
the photosphere resulted in the next explosion of the recurrent Nova. 
In the case of a freely expanding shell, our model predicts $\gamma$-ray fluxes, which are above the sensitivity 
limit of the CTA, at the time shorter than $\sim 1$ yr (for optimum parameters of the shell).
 
The $\gamma$-ray fluxes, produced in the model of a freely expanding shell, can be additionally dissolved in time
due to the difference in distances of specific parts of the shell to the observer (the propagation times of 
produced
$\gamma$-rays may significantly differ). This time delay is of the order of 
$t_{\rm dis} = R_{\rm sh}/c = t_{\rm sh}v_{\rm sh}/c\approx 3.65 t_{\rm yr}v_8$~days, where 
$t_{\rm sh} = 3\times 10^7t_{\rm yr}$~s is the propagation time of the shell. This effect is important in 
the case of the $\gamma$-ray emission
from the shell at the moment of the  recurrence time of the Nova, when the radiation from the photosphere 
of the next RS Oph Nova explosion dominate over the radiation from the Red Giant. 

Although, we took into account energy losses of electrons on different radiation processes, we calculated the 
$\gamma$-ray spectra only from the Inverse Compton process. In fact, other radiation processes can even dominate 
as the main energy loss process of electrons with specific  energies or at specific time after explosion. 
They are not expected to contribute significantly to the 
GeV-TeV $\gamma$-ray energy range. For example, the bremsstrahlung process is important only for electrons with 
energies below a few GeV, but  only at a few days after explosion (see Fig.~2). On the other hand, the synchrotron 
process becomes the main energy loss process for electrons at the largest (multi-TeV) energies at early time 
after explosion, when the maximum 
energies of electrons are due to the synchrotron energy losses (see Figs.~2 and 3). 
At late time after Nova explosion, the maximum energies of electrons are due to the balance between the time scale 
for the electron acceleration and the dynamical time scale of the expending shell (Fig.~3).
Then, IC energy losses of electrons dominate over the energy losses on the synchrotron process (see Fig.~2).
For typical energies of electrons, 
$E = 1E_{\rm TeV}$~TeV, and the strengths of the magnetic field, $B = 10^{-3}B_{\rm mG}$~G,  
at 2 yrs after Nova explosion (see Eq.~2), the synchrotron emission falls into the UV to soft X-ray energy range, 
$\varepsilon_{\rm syn}\approx m_{\rm e}c^2(B/B_{\rm cr})\gamma_{\rm e}^2\approx 46B_{\rm mG}E_{\rm TeV}^2$~eV, $B_{\rm cr} = 4.4\times 10^{13}$ G is the critical magnetic field strength.
This radiation can be over-come by either the hard thermal X-ray emission, with the peak luminosity of the order of 
$\sim 10^{33} - 10^{34}$ erg~s$^{-1}$, produced by the hot gas heated by the shocks within the shell
(Mukai et al.~2008, Gordon et al.~2021), or by the super-soft X-ray emission from the WD surface 
(Kahabka \& van den Heuvel~1997). 

In the considered here two-component geometrical model, we do not take into account the modification of 
the electron spectrum due to the adiabatic expansion (or contraction) of the shell. We neglected this effect since 
it is difficult to predict how the shell behaves during its propagation in the medium surrounding the recurrent Nova. 
The shell might naturally expend due to internal pressure, but it might also contract due to the external 
pressure of
either the surrounding interstellar medium, or the wind from the White Dwarf.
Therefore, we leave this complicated problem for the future more detailed modelling. 

In our modelling, we use simple prescription for the magnetic field during the propagation of the Nova shell. In fact,
if the surrounding of the shell is strongly inhomogeneous, then the downstream turbulent magnetic field would grow 
to milliGauss (see e.g. Giacalone \& Jokipii 2007, Inoue et al. 2012; Fraschetti 2013), which can be above the values predicted by our simple prescription in the case of low value of the magnetization parameter $\alpha$ at a few years after the Nova explosion. Then, the faster 
synchrotron cooling might prevent acceleration of electrons to TeV energies (Vink \& Laming 2003).
As a result, the $\gamma$-ray spectra, predicted for low values of $\alpha$ (and years after explosion), 
might be tested only in the GeV energy range by the satellite observatories.

\section*{Acknowledgments}
I would like to thank the Referee for useful comments.
This work is supported by the grant through the Polish National Research Centre  No. 2019/33/B/ST9/01904.

\section*{Data Availability}
The simulated data underlying this article will be shared on
reasonable request to the corresponding author.


\label{lastpage}

\end{document}